\documentclass[sn-mathphys,Numbered]{sn-jnl}

\usepackage{graphicx}%
\usepackage{multirow}%
\usepackage{amsmath,amssymb,amsfonts}%
\usepackage{amsthm}%
\usepackage{mathrsfs}%
\usepackage[title]{appendix}%
\usepackage[dvipsnames]{xcolor}%
\usepackage{textcomp}%
\usepackage{manyfoot}%
\usepackage{booktabs}%
\usepackage{algorithm}%
\usepackage{algorithmicx}%
\usepackage{algpseudocode}%
\usepackage{listings}%
\usepackage{verbatim}
\usepackage{braket}
\usepackage{ulem}
\usepackage{pdfpages}
\usepackage{import}
\usepackage{soul}

\newcommand{\B}{\mathbf}

\begin{document}

\title[Article Title]{Signatures of Mode-Resolved, Nonlocal Electron-Phonon Coupling in Two-Dimensional Spectroscopy}

\author[1]{\fnm{Sheng} \sur{Qu}}

\author[2]{\fnm{Vishal K.} \sur{Sharma}}

\author[1,3]{\fnm{Jaco J.} \sur{Geuchies}}

\author[1]{\fnm{Maksim} \sur{Grechko}}

\author[1]{\fnm{Mischa} \sur{Bonn}}

\author*[2]{\fnm{Falko} \sur{Pientka}}

\author*[1,4]{\fnm{Heejae} \sur{Kim}}\email{heejaekim@postech.ac.kr}
\email{pientka@itp.uni-frankfurt.de}

\affil[1]{\orgdiv{Department of Molecular Spectroscopy}, \orgname{Max-Planck Institute for Polymer Research}, \orgaddress{\city{Mainz}, \postcode{55128}, \country{Germany}}}

\affil[2]{\orgdiv{Institut f\"ur Theoretische Physik}, \orgname{Goethe-Universit\"at}, \orgaddress{\city{Frankfurt a.M.}, \postcode{60438}, \country{Germany}}}

\affil[3]{\orgdiv{Leiden Institute of Chemistry}, \orgname{Leiden University}, \orgaddress{\city{Leiden}, \postcode{2333CC}, \country{Netherlands}}}

\affil[4]{\orgdiv{Department of Physics}, \orgname{Pohang University of Science and Technology}, \orgaddress{\city{Pohang}, \postcode{37673}, \country{Korea}}}

\abstract{Electron-phonon coupling (EPC) is foundational in condensed matter physics, determining intriguing phenomena and properties in both conventional and quantum materials. In this manuscript, we propose and demonstrate a novel two dimensional (2D) EPC spectroscopy which allows for direct extraction of EPC matrix elements for specific phonon modes and different electron energies, simultaneously. Using this technique, we are able to measure the electron-energy dependence of the EPC strength for individual phonon modes. This capability allows us to identify unique signatures distinguishing nonlocal Su-Schrieffer-Heeger (SSH) -type couplings from local Holstein-type couplings. In application to a methylammonium lead iodide (MAPI) perovskite, we find that two pronounced phonon modes at room temperature exhibit highly distinctive EPC behaviors, concerning strength, anisotropy, and temperature-dependence across the structural phase transition. Our approach paves the way for unraveling the microscopic origin of EPC, the change of the phonon-mode-specific EPC  with external conditions, and phonon-mediated ultrafast control of condensed materials.}


\maketitle

\section{Introduction}\label{sec1}

Interactions among constituent particles in many-body systems govern the physics of condensed matter systems. Electron-phonon couplings (EPCs), in particular, affect the temperature dependence of carrier mobility and resistance, the electronic energy band structures, the phonon dispersion relations, and the lifetimes of those particles in solids \cite{Giustino_2017}. The roles EPCs play in ordered phases, such as conventional superconductivity, charge-density wave order, and metal-insulator transitions are the subject of vast research fields \cite{PhysRevX.6.041019}. 
Understanding the role of phonons in high-temperature superconductors \cite{Lanzara_2001} requires a careful modeling of the microscopic nature of EPCs, including nonlocal effects that lead to a variation of EPCs with electron momentum $\B k$ \cite{Devereaux2004}. In fact, a strong coupling of phonons to the kinetic energy of the electrons rather than the potential has been predicted to lead to considerably higher critical temperatures for phonon-mediated superconductors \cite{Sous2018,Cai2023}. It is therefore crucial to identify materials that go beyond the paradigm of $\B k$-independent EPC in the Holstein or Fr\"ohlich models \cite{mahan2013many}. Moreover, via specific phonon modes, one could modulate band structures \cite{Rini_2007,Subedi_2014,Hu_2014,Kim_2017}, transport properties \cite{Sekiguchi_2021,Lan_2019,Leitenstorfer_1999}, and topology \cite{Hasan_2010,Weber_2018} on (sub-)picosecond time scales and with high precision in atomic motions. Such an ultrafast phonon control  allows for achieving electronic states not accessible in static materials \cite{Li_2019,Waldecker_2017}. 

In order to advance our fundamental understanding of emergent phenomena and capability to exploit new functionalities, it is essential to differentiate contributions from individual phonon modes and different electronic energies to the overall EPC strength. Most experimental approaches, however, provide rather indirect information on EPCs without an explicit phonon-mode-resolution. Conventional techniques, such as transport measurements, heat capacity or angular-resolved photoelectron spectroscopy (ARPES) measure the renormalization of response functions or the band structure due to phonons. Such quantities typically include the average effect of many phonon modes and are also subject to other phenomena such as electron-electron interaction, rendering the phonon-mode-resolved EPC strength extraction impossible or unreliable \cite{Calandra_2007,Park2008}. While inelastic X-ray scattering \cite{Ament_2011,Mohr_2007} and time-resolved ARPES have been proposed \cite{De_Giovannini_2020,Calandra_2007} for this purpose, these approaches have practical limitations coming from their intrinsic surface sensitivity, resolution at low energies and additional material specific requirements \cite{Ament_2011,Na_2019,Lanzara_2001,Lee_2014}. 

Here, we develop a two-dimensional (2D) spectroscopy protocol that allows for direct extraction of EPC matrix elements for individual optical phonon modes and different electron energies (or electron momenta for a known dispersion), simultaneously. In this 2D electron-phonon coupling spectroscopy (2D EPC), the EPC strengths of a solid manifest themselves as signal peaks in a measured 2D spectrum, where each of two axes corresponds to the zone-center phonon frequencies and the electron energies, respectively. We demonstrate how the EPC matrix elements are extracted from the measured 2D EPC signal without requiring extensive modeling. The interpretation of the measurement is facilitated by the fact, that the signal vanishes in the absence of EPC. This is a direct manifestation of a general principle in 2D spectroscopy: cross peaks can only arise because of correlations between excitation and detection frequencies \cite{hamm_2011}. Since our approach is all-optical and bulk sensitive, it is applicable to almost every kind of gapped solid. It is distinctly different from techniques that rely on phonon-induced relaxation of electronic states like the recently demonstrated time- and angle-resolved photoemission experiments \cite{Na_2019}, which are limited to electronic states with a restricted scattering phase space and with a sufficiently strong coupling to phonons. In contrast, our signal exclusively arises from electron-phonon correlations, making our experiment sensitive even to weak couplings while avoiding any averaging over electronic states.

To showcase our technique, we measure the EPC strength of two pronounced phonon modes in methylammonium lead iodide perovskite (MAPI) as a function of electron energy. We find a surprisingly strong dependence on electron energy, including a vanishing of the EPC strength at the electronic $\Gamma$ point. This result indicates a pronounced coupling of phonons to the electronic kinetic energy in MAPI akin to an SSH-type EPC. Hence, our technique is able to distinguish different microscopic EPC models by detecting the characteristic electron momentum dependence of the EPC strength, which is beyond the reach of conventional methods. In addition, we perform temperature and polarization dependent measurements to highlight the potential of our 2D EPC technique for studying the evolution of mode-resolved EPC as a function of various phase space parameters (e.g. temperature, pressure, strain, doping, and chemical composition). 

\section{Results}\label{sec2}

The basic concept of 2D EPC spectroscopy is illustrated in the schematic diagram in Fig. 1(a), where a series of three laser pulses interacts with the material. These interactions induce a polarization whose contribution at third order in the electric field of the light waves is subsequently measured. To generate correlated excitations of specific electronic states and phonon modes, we use one pulse with broadband THz frequency (denoted by THz in Fig. 1(a)) and two with optical frequencies. The individual frequencies of these two optical pulses are chosen to be non-resonant with any elementary excitation, while their sum is resonant with an electronic interband transition as illustrated in the energy diagram in Fig. 1(b).
One of the two optical pulses has a broad bandwidth (denoted by BB), and the other a narrow bandwidth (denoted by NB), so that the NB spectrum can be approximated by a delta-function. (for its detailed use in analysis, see Methods). 

Figure 1(c) displays the time evolution of the initial state (denoted by a density matrix $\ket{a}\bra{a}$) subject to successive radiation-matter interactions (colored arrows). The final state can then be obtained by summing over all possible Liouville space pathways \cite{mukamel_1999} shown in Fig. 1(c). The radiation-matter interaction is described by the semiclassical Hamiltonian under the dipole approximation as $H_\mathrm{int}(t')=-E(\B r,t')\cdot \mu$, where $E(\B r,t')$ is the position and time-dependent electric field, and $\mu$ is the dipole operator.The induced polarization at third order in the electric field, $P^{(3)}(\B r, t')$, is determined by the expectation value of the third-order density matrix, $\rho^{(3)}(t')$, and can be expressed in terms of the third-order response function, $S_{\rm EPC}(t_3,t_2,t_1)$ (see Methods and Ref.~\cite{mukamel_1999}).After a 2D Fourier transformation over times $t$ and $\tau$ defined in Fig.~1(a) (denoting the Fourier frequencies by $\omega_{t/\tau}$), the 2D EPC signal field, $E_{\rm EPC}(\omega_t,\omega_\tau)$, is obtained in terms of the response function, $S_{\rm EPC}(\omega_t,\omega_\tau)$, and the electric fields of the incident THz (BB) pulses, $E_{\rm THz}$ ($E_{\rm BB}$) (see Methods):
\begin{equation} 
    E_{\rm EPC}(\omega_t,\omega_\tau) \propto  S_{\rm EPC}(\omega_t,\omega_\tau) E_{\rm THz}(\omega_\tau)E_{\rm BB}(\omega_t-\omega_{\rm NB}-\omega_\tau).
\end{equation}
We will see below that the the response function $S_{\rm EPC}(\omega_t,\omega_\tau)$ is a measure of the coupling strength of phonons and electronic excitations at frequencies $\omega_\tau$ and $\omega_t$, respectively.

The starting point of our theoretical description of the material is the Hamiltonian $H=H_0+H_{\rm ep}$, which contains 
a noninteracting part $ H_{0}=\sum_{k, n} \varepsilon_{n,\mathbf{k}} {c}^{\dagger}_{n,\mathbf{k}} {c}_{n, \mathbf{k}} +  \sum_{\mathbf{q}, \lambda}   \omega_{\mathbf{q}, \lambda} \left({b}^{\dagger}_{\mathbf{q}, \lambda} {b}_{\mathbf{q}, \lambda} +1/2  \right)$, as well as a generic electron-phonon interaction to first order in nuclear displacements \cite{mahan2013many,Giustino_2017},
\begin{equation}\label{e-p coupling}
   H_{\mathrm{ep}} =   \sum_{\mathbf{k},\mathbf{q}, n, \lambda}  \omega_{\mathbf{q}, \lambda} M_{\mathbf{k, q}, n, \lambda} {c}^{\dagger}_{n ,\mathbf{k+q}} {c}_{n, \mathbf{k}} \left({b}_{ \mathbf{q}, \lambda} +{b}^{\dagger}_{-\mathbf{q}, \lambda}  \right).
\end{equation}  
Here,  $M_{\mathbf{k, q}, n, \lambda}$ is the dimensionless coupling strength, the operator ${c}_{n,\mathbf{k}}$ (${c}^{\dagger}_{n,\mathbf{k}}$) annihilates (creates) an electron with band index $n$ and energy $\varepsilon_{n,\mathbf{k}}$, the operator ${b}_{\mathbf{q}, \lambda} ({b}^{\dagger}_{\mathbf{q}, \lambda})$ annihilates (creates) a phonon of mode $\lambda$ with occupation number $v$ and energy $ \omega_{\mathbf{q}, \lambda}$, and we have set $\hbar=1$. We specifically consider an electronic two-band model with $n=g,e$, labeling the ground- and excited-state bands. The energy level diagram of this model at fixed electron momentum $\mathbf{k}$ and $\mathbf{q}=0$ corresponding to the single-particle eigenstates $| n_{\mathbf{k}},\chi_{n,\lambda}^v \rangle$ is shown in Fig. 1(b) using the short-hand notation $\ket{n,v}\equiv \ket{n_{\mathbf{k}},\chi_{n,\lambda}^v}$.

\vspace{1cm}
\noindent\includegraphics[width=\textwidth]{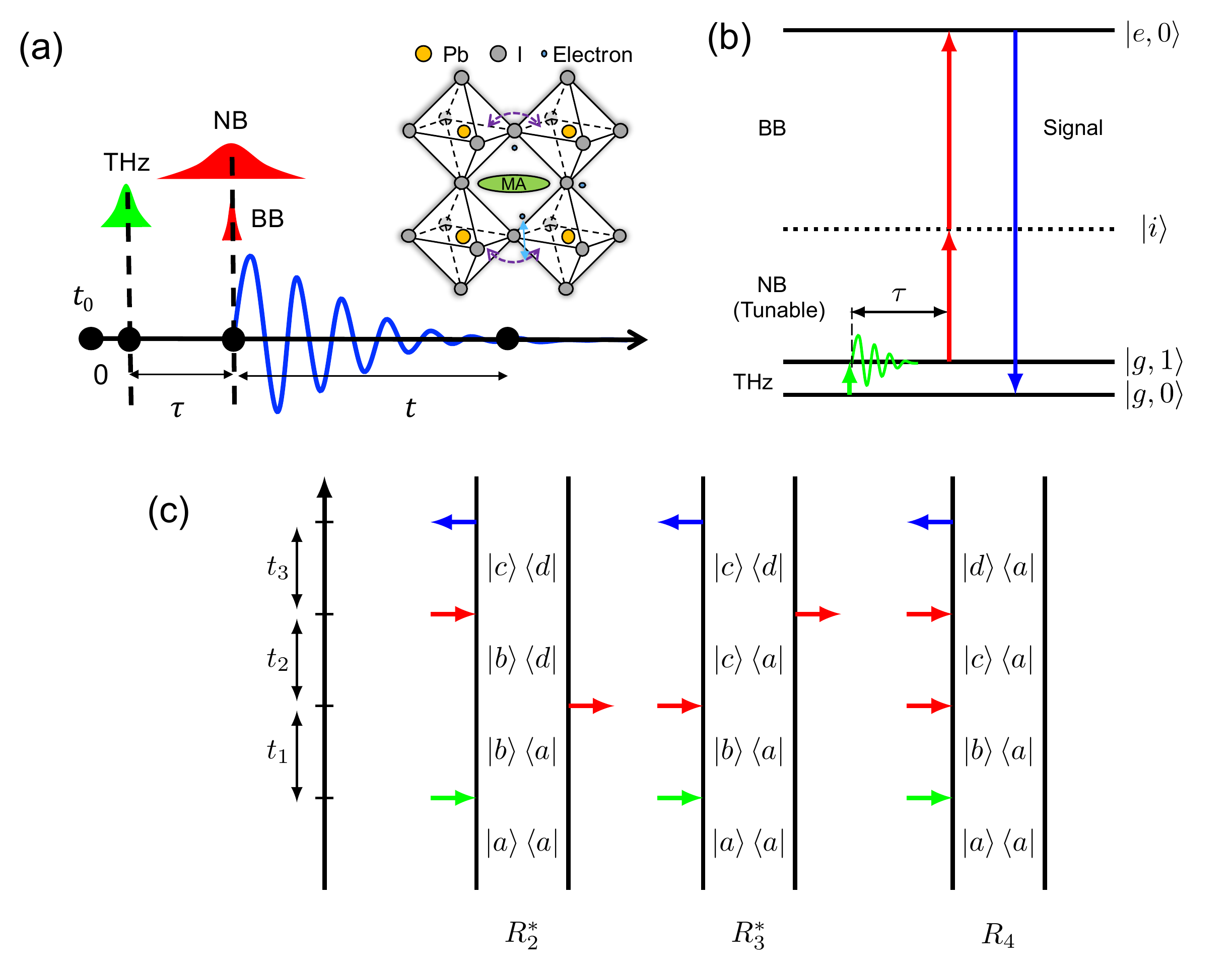}\\
\noindent {\bf Figure 1. Schematic Representations of 2D EPC Spectroscopy} (a) The pulse configuration for the experimental 2D EPC spectroscopy with spectral characteristics suited for our target material, MAPI. The black circles denote the time-ordered points when the interactions between the radiation field and matter take place and $\tau$ and $t$ are the time intervals between these interactions.  (b) Energy level diagram of a noninteracting electron-phonon system at a given electron momentum $\B{k}$ and phonon-momentum $\B{q}=0$ corresponding to the single-particle eigenstates $\ket{n,v}\equiv\ket{n_{\B{k}},\chi^{v}_{n,\lambda}}$. Here, $|i\rangle$ denotes a nonresonant intermediate state, which is one of the eigenstates. The green and two red colored vertical arrows indicate interactions with THz, NB and BB NIR pulses in (a), respectively. The blue downward arrow denotes the signal field emission. (c) The Liouville space pathways contributing to the third-order response function in Eq. (\ref{resp_func}). The vertical double lines represent the \textit{ket} and \textit{bra} side of the density operators that evolve in time along the upward direction. The colored horizontal arrows represent interactions with the radiation fields of the correspondingly colored pulses in (a). The time intervals between these interactions are denoted by $t_1$, $t_2$, and $t_3$.
\vspace{1cm}

Then, we proceed by absorbing the EPC term of the Hamiltonian into a dressed phonon, redefining the phonon operator $b_{\B q,\lambda}\to B_{\B q,\lambda}$ as
\begin{equation} \label{redef_phonon}
B_{\B q,\lambda}=b_{\B q,\lambda}+\sum_{\B k,n}M^*_{\B k,\B q,n,\lambda}c^\dag_{n, \B k}c_{n, \B k+\B q}.
\end{equation}
The Hamiltonian becomes diagonal in the transformed phonon basis, $H=\sum_{\mathbf{q}, \lambda}   \omega_{\mathbf{q}, \lambda} \Bigl({B}^{\dagger}_{\mathbf{q}, \lambda} {B}_{\mathbf{q}, \lambda} +\frac{1}{2}  \Bigr)+H_{\rm el}$, where the electronic part of the Hamiltonian $H_{\rm el}$ now also contains phonon mediated electron-electron interactions and we have used the identity $M^*_{\B k+\B q,-\B q,n,\lambda}=M_{\B k,\B q,n,\lambda}$ which follows from the hermiticity of Eq.~(\ref{e-p coupling}).
We denote electrons in the many-body ground state (i.e., forming a Fermi sea) by $\ket{g}$; and those in the optically excited states with a single electronic interband transition at momentum $\B k$ by $\ket{e_\B k}=c_{e,\B k}^\dag c_{g,\B k}\ket{g}$. In the following, we drop the index $\mathbf{q}$, focusing on the $\mathbf{q}=0$ sector relevant to this optical approach. In this sector, the dressed phonon operator $B_{\B q=0,\lambda}$ is simply the original phonon operator shifted by a constant that depends on the occupation of the electronic bands. We will henceforth refer to this shifted operator as the ``phonon" with the implication that its eigenstates depend on the electronic state. For instance, the phonon vacuum at $\B q=0$  for the electronic ground state is defined by
\begin{align}
{B}_{\lambda}\ket{g,\chi^{v=0}_{g,\lambda}}=\Bigl(b_{\lambda}+\sum_{\B k'}M^*_{\B k',g,\lambda}\Bigr)\ket{g,\chi^{v=0}_{g,\lambda}} =0. 
\end{align}
In contrast, the phonon vacuum for an electronic excited state $\ket{e_\B k}$ is defined by
\begin{align}
{B}_{\lambda}\ket{e_\B k,\chi^{v=0}_{e,\B k,\lambda}}=\Bigl(b_{\lambda}+\sum_{\B k'}M^*_{\B k',g,\lambda}+M^*_{\B k,e,\lambda}-M^*_{\B k,g,\lambda}\Bigr)\ket{e_\B k,\chi^{v=0}_{e,\B k,\lambda}} =0
\end{align}
which means that $\ket{\chi^{v=0}_{e,\B k,\lambda}}$ takes the form of a coherent state when expressed in the basis $\ket{\chi^{v=0}_{g,\lambda}}$.

The phonon-mode- and electron-momentum-dependent EPC strengths, $M_{\mathbf{k}, n, \lambda}$, can be associated with the the measured response function, $S_{\rm EPC}(t_3,t_2,t_1)$. To this end, we consider all possible Liouville space pathways representing different  time orderings of overlapping pulses and different intermediate states of the multiphoton process (see SI).   The magnitude of the response function is proportional to all transition matrix elements, $\mu_{n'v',nv} = \bra{n'_{\mathbf{k}},\chi_{n',\lambda}^{v'}} \mu \ket{n_{\mathbf{k}},\chi_{n,\lambda}^{v}}$, involved in a given pathway. 
For the ``nonrephasing" contribution, where the phase of the induced polarization is acquired continuously during $t_1$ and $t_3$,  the three diagrams $R_2^*$, $R_3^*$, and $R_4$ shown in Fig. 1(c)  need to be taken into account (see  Fig. S1 and Ref.~\cite{mukamel_1999}). The nonlinear response function is given by the sum over all possible pathways (meaning all possible diagrams and intermediate states), and we find
\begin{align}
    S_{\rm EPC}(t_3,t_2,t_1)&\propto -\sum^{R_2^*}_{abcd}m_{abcd}
    e^{-it_1\omega_{ba}-it_2 \omega_{bd}-it_3 \omega_{cd}}
    - \sum^{R_3^*}_{abcd}m_{abcd}
    e^{-it_1\omega_{ba}-it_2 \omega_{ca}-it_3 \omega_{cd}}\notag\\
&    
+\sum^{R_4}_{abcd}m_{abcd}e^{-it_1\omega_{ba}-it_2 \omega_{ca}-it_3 \omega_{da}},\label{resp_func}
\end{align}
where the three sums only run over states $\ket{a}=\ket{n,v}$ allowed by the respective diagrams in Fig. 1(c). Here, we use the short-hand notation $m_{abcd}=\mu_{ba}\mu_{ad}\mu_{cb}\mu_{dc}$ and $\omega_{ba}$ denotes the energy difference between states $\ket{b}$ and $\ket{a}$. 

In the absence of EPC, the transition matrix elements are simply  $
\mu_{n'v',nv} = \delta_{v'v}\bra{n'_{\mathbf{k}}} \mu_{e} \ket{n_{\mathbf{k}}} + \delta_{n'n}\bra{\chi_{n,\lambda}^{v'}} \mu_{p} \ket{\chi_{n,\lambda}^v}$ with $\bra{\chi_{n,\lambda}^{v'}} \mu_{p} \ket{\chi_{n,\lambda}^v} \propto \delta_{v',v\pm 1}$. Hence a single photon either leaves  the phonon number  unchanged (interband transition) or changes it by $\pm 1$ (intraband transition).
The first photon in Fig. 1(c) resonantly excites a single phonon and so we can fix $\ket{a}=\ket{g,0}$ and $\ket{b}=\ket{g,1}$ for all three diagrams. The second and third photon do not need to be resonant, but if we focus on the nonrephasing contribution (\textit{i.e.}, the same sign between $\omega_\tau$ and $\omega_t$), we only find one possible set of intermediate states, $\ket{c}=\ket{e,1}$ and $\ket{d}=\ket{g,1}$, for the first two diagrams. For the third diagram, there are two possibilities with $\ket{c}=\ket{g,0}$ or $\ket{c}=\ket{e,1}$ and $\ket{d}=\ket{e,0}$. The detailed diagrams for both nonrephasing and rephasing contributions at all possible phase matching directions are provided in the Supplementary Fig. S1. In the absence of EPC, the electronic transitions do not depend on the phonon occupation and vice versa, i.e., $\mu_{g0,e0}=\mu_{g1,e1}$ and $\mu_{g0,g1}=\mu_{e0,e1}$. We can then readily convince ourselves that the product of matrix elements in Eq.~(\ref{resp_func}) is identical for all three diagrams. Evaluating the sum over the four remaining terms yields
\begin{align}
    S_{\rm EPC}(t_3,t_2,t_1)\propto |\mu_{g1,g0}|^2|\mu_{e0,g0}|^2  e^{-it_1\omega_{g1,g0}}
(&-e^{-it_3 \omega_{e1,g1}}
-e^{-it_2 \omega_{e1,g0}-it_3 \omega_{e1,g1}}\notag\\
&+e^{-it_3 \omega_{e0,g0}}
+e^{-it_2 \omega_{e1,g0}-it_3 \omega_{e0,g0}}
)
\end{align}
which equals zero. The response function considering all the rephasing contributions also undergoes the total cancellation in a similar manner. This means the response function and therefore the signal vanishes when the phonons do not couple to the electronic transition. This cancellation has the same origin as the vanishing of the cross peaks in the absence of, \textit{e.g.}, intermolecular interactions in 2D IR spectroscopy \cite{mukamel2000}.

The presence of EPC, on the other hand, has two important consequences for the 2D EPC spectrum. First, the total spectroscopic signal is nonzero and, in fact, a function of EPC strength as it depends on the difference of matrix elements $\mu_{e0,g0}-\mu_{e1,g1} \propto M^{2}_{\mathbf{k},\lambda}e^{-M^{2}_{\mathbf{k},\lambda}/2}$, with the effective EPC strength  $M_{\mathbf{k}, \lambda} \equiv M_{\mathbf{k}, e,\lambda}-M_{\mathbf{k},g, \lambda}$. Therefore, while the intraband phonon transitions are independent of the EPC, the interband transitions only depend on the effective EPC strength, $M_{\mathbf{k}, \lambda}$. The fact that the 2d EPC spectrum is sensitive to the difference of EPC strength from conduction and valence band is a common feature of experimental methods based on optical transitions \cite{Patrick_2014}. Second, part of the spectral weight of the interband transitions is transferred to higher phonon numbers and so additional pathways start to contribute (see Discussion). It is noteworthy that EPC can also result in a phonon frequency shift with band index, which is not included here. However, we believe our results below to remain qualitatively true in that case as well, as such effects will similarly destroy the perfect cancellation explained above.

\vspace{1cm}
\noindent\includegraphics[width=.92\textwidth]{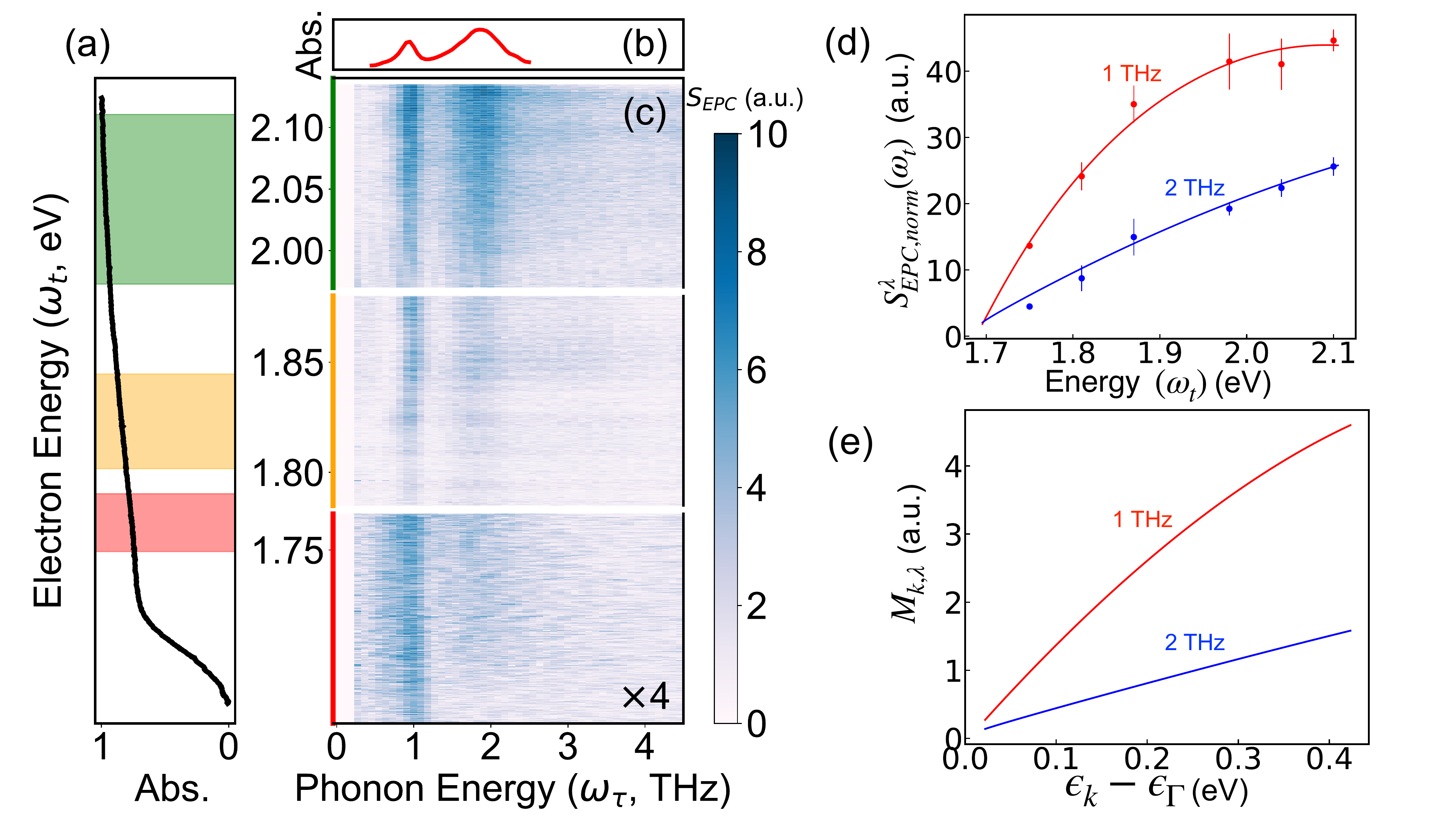}
\noindent {\bf Figure 2. Experimental 2D EPC Spectrum} (a) The linear optical absorption spectrum, (b) the THz absorption spectrum, and (c) the 2D EPC spectrum, $S_{\rm EPC}(\omega_t, \omega_\tau)$, along the electron and phonon energies for MAPI measured at room temperature. (d) The normalized experimental 2D EPC response function for the given phonon mode, $S_{\mathrm{EPC,norm}}^{\lambda} (\omega_t)$,  with the phonon modes denoted by $\lambda=1, 2$ THz.  The closed circles indicate the experimental values, averaged over the electronic (phonon) energy range of $\sim$ 0.01 eV ($\sim$ 0.1 THz), and the solid lines the theoretical result obtained from Eq.~(\ref{Ptot}). The error bar indicates the standard deviation of $>$~100 data points over the range. (e) Extracted relative electron-phonon coupling strength $M_{\mathbf{k}}$ as a function of the electronic transition energy $\varepsilon_\B k = \varepsilon_\Gamma + \Delta \cos (ka)$ for both phonon modes.
\vspace{1cm}

For the experimental demonstration, the pulse characteristics are chosen to match the energy scales of the MAPI at room temperature. The THz pulse with a spectral range up to 60 meV contains a sufficient coverage to excite all inorganic-sublattice-based phonon modes reported in the MAPI (Fig. S2(a)) \cite{Leguy_2016}. In order to gain access to the broad range of electronic transition energies above the band gap of MAPI, we tune the NB pulse from 1300 to 2000 nm and produce a BB pulse with a spectral width of 1000 - 1250 nm (Fig. S2(b)). Both of these photon energies are away from any elementary excitations in MAPI. These two optical (i.e. BB and NB) pulses collinearly propagate through the sample simultaneously, and the delay time between the optical pulse pair and the THz pulse, $\tau$, is scanned. The phase of the signal field, $E_{\rm EPC}(t,\tau)$, is resolved by interference with a local oscillator (LO) field (Fig. S2(c)). The LO field is generated after the sample, and both fields are dispersed at a spectrograph and detected by a camera.  The thus measured intensity $I_{\rm EPC}(\omega_t,\tau)$ containing the squared magnitude of the total field, $E_{\rm tot}(t,\tau)=E_{\rm EPC}(t,\tau)+E_{\rm LO}(t) + c.c.$, is then Fourier-transformed along the delay time $\tau$, producing the frequency domain 2D EPC spectrum,  $I_{\rm EPC}(\omega_t,\omega_\tau)$ (Fig. S3, see Methods). 

From this measured spectrum, $I_{\rm EPC}(\omega_t,\omega_\tau)$, we extract the 2D EPC response function, $S_{\rm EPC} (\omega_t,\omega_\tau)$, by  the following analysis protocol. First, we isolate the desired phase matching condition. To do so, we introduce an additional time delay between the LO and the signal field that contributes to the opposite responses between the rephasing and nonrephasing pathways, and separate it temporally (Eq.~\eqref{S9}-\eqref{S10} and Fig. S4-S5). Second, we correct the relative phase differences among three incident pulses at the sample by using a SiN membrane as a reference sample. We thus obtain the real and imaginary spectra for both nonrephasing and rephasing contributions, separately (Fig. S6). In fact, such a full separation could be useful for analyzing the lineshape of isolated resonances \cite{Li2006}.  Here, we focus on the slightly larger nonrephasing contribution. Finally, we normalize the isolated 2D spectrum, $I_{\rm EPC,\mathbf{k}_s}(\omega_t,\omega_\tau)$, by the spectra of $E_{\rm THz}(\omega_\tau)$ and $E_{\rm LO}(\omega_t)$. The LO spectrum is, in fact, the upconverted BB spectrum by the NB frequency (Fig. S2(c), see Methods). 

The nonrephasing 2D EPC spectrum of MAPI obtained from this analysis, $S_{\rm EPC} (\omega_t,\omega_\tau)$, is shown in Fig. 2(a) for a range of electron energies (1.6 $\sim$ 2.2 eV). Evidently, both 1 and 2 THz modes couple to electrons with energy above the band gap to some extent as they contribute to the 2D EPC signal. The signals from both modes monotonically increase as the electron energy increases up to 0.6 eV above the band gap, with the contribution of the 1 THz mode consistently larger than  that of the 2 THz mode.

For quantifying the EPC strength at a given phonon mode, $\lambda$, and electronic energy $\varepsilon_{\mathbf{k}}$, it is important to remember that the 2D EPC spectrum, $S_{\rm EPC} (\omega_t,\omega_\tau)$ (shown in Fig. 2(a)) depends on four transition matrix elements, namely, two phononic and two electronic transition elements (\textit{e.g.}, in the top right diagram of Fig. 3(a) we have $\mu_{g1,g0}$, $\mu_{em,e(m+1)}$ for the phononic and $\mu_{e(m+1),g1}$, $\mu_{g0,em}$  for the electronic transitions). Moreover, the 2D EPC signal field contains the phonon and electron density of states. Therefore, in order to isolate the pure contribution of the EPC strength to the 2D EPC response function for a given phonon mode $\lambda$,  we normalize the measured response function, $S_{\rm EPC} (\omega_t,\omega_\tau)$, (Fig. 2(a)) by both absorbance spectra along the electronic (Fig. 2(b)) and phononic (Fig. 2(c)) energies, resulting in $S_{\mathrm{EPC,norm}}^{\lambda} (\omega_t)$ (closed circles in Fig. 2(d), for theoretical analysis see Discussion). We note here that our approach is applicable to generic nuclear displacements, but the quantitative comparison of EPC strengths can only be performed for phonon modes with non-zero oscillator strengths. The result indicates a consistently stronger coupling of electrons in this energy range to the 1 THz mode than to the 2 THz mode.
Both normalized signals monotonically increase from zero near the band gap and the growth slows down as the electron energy approaches 2 eV.

\section{Discussion}\label{sec3}
Using our theoretical model, we can relate the measured 2D EPC signal to the EPC strength, $M_{\mathbf{k}, \lambda}$. Our starting point is an expression for the induced polarization, $P_{\rm EPC,\mathbf{k}}(\omega_t,\omega_\tau)$ at fixed electronic momentum, $\mathbf{k}$ (see Methods). Because transitions can involve multiple phonon excitations in the presence of EPC, the signal field is emitted at a series of discrete frequencies spaced by the phonon frequency, $\varepsilon_\mathbf{k}+m \omega_{\lambda}$, where $\varepsilon_{\mathbf{k}}=\varepsilon_{e,\mathbf{k}}-\varepsilon_{g,\mathbf{k}}$ is the momentum dependent electronic transition energy. Considering the non-rephasing contribution (i.e., $\omega_\tau > 0, \omega_t > 0$), such emissions take place via four different pathways shown in  Fig. 3(a) We evaluate these diagrams explicitly in the Supplementary Material and the resulting induced polarization is given by $P_{\rm EPC,\mathbf{k}}(\omega_t,\omega_\tau)\propto  \sum_m \delta(\epsilon_\B k+m\omega_\lambda-\omega_t)C_{m,\B k}e^{-M_{\B k,\lambda}^{2}}(M_{\B k,\lambda})^{2m}(M_{\B k,\lambda}^2-m)/m!$, which is shown in Fig. 3(b). Here, $C_{m,\B k}$ is a constant that weakly depends on $\B k$ (see Methods). 
The polarization changes sign as a function of the detection frequency, $\omega_t$, crossing zero at a frequency corresponding to the typical number of additional phonon excitations, $m=M_{\mathbf{k}, \lambda}^2$. Interestingly, the sum over all frequencies, $\int d\omega_t P_{\rm EPC,\mathbf{k}}(\omega_t)$, vanishes regardless of the EPC strength. This cancellation originates from the destructively interfering pathways discussed above. However, in the presence of EPC the negative contributions to the induced polarization acquire a stronger blueshift than the positive ones leading to an imperfect cancellation. This leads to the characteristic dispersive lineshape in Fig.~3(b) expected for 2d spectra of discrete levels \cite{Li2006}. Our measured spectrum, however, has a different lineshape as the electronic excitations form a continuum and the full signal involves a convolution of the density of states with  $P_{\rm EPC,\mathbf{k}}$.

The total induced polarization at a particular frequency  $\omega_t$ is given by a sum over all electronic momenta 
\begin{align}
 P_{\rm EPC, tot}(\omega_t,\omega_\tau)=\int d\varepsilon_\B k \nu(\varepsilon_\mathbf{k}) P_{\rm EPC, \mathbf{k}}(\omega_t,\omega_\tau).\label{Ptot}
\end{align}
where $\nu$ is the density of states corresponding to the transition energy  $\varepsilon_{\mathbf{k}}$. The density of state is nonzero for arguments above some minimal energy $\varepsilon_\Gamma$, which is the smallest energy separation between electronic ground and excited states with the same momentum. Equation (\ref{Ptot}) can be used to numerically evaluate the 2D EPC spectrum for a given momentum-dependent coupling strength $M_{\B k,\lambda}$. In order to determine the EPC strength from a measured spectrum, we use an analytical approximation. Specifically, we assume the coupling strength  $M_{\varepsilon_{\mathbf{k}},\lambda}\equiv M_{\mathbf{k},\lambda}$  to vary little with $\varepsilon_{\mathbf{k}}$ on the scale of the phonon frequency $\omega_{\lambda}$, and we obtain
\begin{align}
 P_{\rm EPC,tot}(\omega_t,\omega_\tau)\propto  f(\omega_\tau)\frac{d}{d\omega_t}[C(\varepsilon_\B k)\nu(\varepsilon_\B k)M_{\varepsilon_\B k,\lambda}^2]_{\varepsilon_\B k=\omega_t}\label{analytics}
\end{align}
where $f(\omega_\tau)= \Gamma \omega_\lambda /[(\omega_\tau-\omega_\lambda)^2+\Gamma^2]$ is the lineshape along $\omega_\tau$ and $C(\varepsilon_\B k)$ is given in the Methods section. This equation can be inverted and we find
\begin{align}
M_{\varepsilon_\B k,\lambda} \propto \Bigl[\frac{1}{\nu(\varepsilon_\B k) C(\varepsilon_\B k)}\int^{\varepsilon_\B k}_0 d\omega_t\nu(\omega_t) S_{\rm EPC,norm}^{\lambda}(\omega_t)\Bigr]^{1/2},\label{invert}
\end{align}
where $S_{\mathrm{EPC,norm}}^{\lambda} (\omega_t)$ corresponds to the normalized 2D EPC spectrum for a given phonon mode (shown in Fig.~2(d)) and we have only omitted prefactors that do not depend on $\omega_\lambda$ and $\omega_t$ (see Methods). 

To determine the phonon-mode- and electron-energy-specific coupling strength from this expression, we fit the data in Fig.~2(d) with a polynomial function for each mode with the same zero crossing and use the transition energy $\varepsilon_{\mathbf{k}} = \varepsilon_\Gamma + \Delta \cos (ka)$. Here, the band gap $\varepsilon_{\Gamma}=1.68\,$eV has been obtained from the zero crossing of the fit functions, and the band width of MAPI, $\Delta=1.2\,$eV was taken from \cite{umari2014relativistic}. The resulting effective EPC strength, $M_{\varepsilon_\B k,\lambda}$, is shown in Fig.~2(e). Note that the intensity of the signal drops to zero at the band edge due to the vanishing density of electronic states below the gap. Independently, we find as a result of our fit that the EPC strength for both modes approaches zero at the band edge. 

With the thus obtained EPC strength, $M_{\varepsilon_\B k,\lambda}$, we can then reproduce the full 2D EPC spectrum. To model the two pronounced phonon modes (namely, the $\sim 1$ THz and $\sim 2$ THz modes), we choose the frequency, width, and oscillator strength to match the experimental THz spectrum in Fig. 2(c). With this we obtain the 2D EPC spectrum shown in Fig. 3(c) by numerically evaluating Eq.$~$(\ref{Ptot}). The calculated normalized spectrum shown as solid lines in Fig.~2(d) is in good agreement with the experimental data.  

The phonon-mode resolved EPC strength as a function of electron energy in Fig.~2(e) is the main result of this paper. We observe a roughly linear drop of the EPC at the band edge $\varepsilon_\B k=\varepsilon_\Gamma$ for both modes, which corresponds to a vanishing of the EPC strength as $\sim k^2$ at $\B k=0$. Such a strong dependence on electron wavevector is in marked contrast to a standard Holstein coupling term, which assumes a coupling of phonons to the onsite density of electrons and does not vary with $\B k$. The behavior in Fig.~2(e) suggests instead an EPC that affects the kinetic energy rather than the potential of the electrons.

To gain further insight, we consider the SSH model as a paradigmatic model with a $\B k$-dependent EPC. In Supplementary Sec. 1.6, we show that the EPC strength for a $\B q=0$ optical phonon mode in the SSH model indeed vanishes at $\B k=0$ as $M_{\B k,\B q=0}\propto k$. The SSH model, however, is not directly applicable to our case as it describes longitudinal phonons. We have therefore developed another toy model for transverse phonons with an EPC that vanishes at $\B q=\B k=0$ (see SI). While the microscopic coupling mechanism in both models is rather distinct, we show that the strong electron momentum dependence of the EPC in the latter can be understood as an effective renormalization of the hopping strengths in the spirit of the SSH model. In contrast to the SSH model, however, the EPC strength  of transverse phonons vanishes quadratically, $M_{\B k,\B q=0}\propto k^2$, as required by symmetry and in agreement with our experimental observations. Moreover, the characteristic behavior at $\B q=\B k=0$ of these two toy models agrees with more realistic electron-phonon models for longitudinal and transverse modes discussed in \cite{Devereaux1995,Devereaux2004} in the context of the cuprates.
We conclude that measuring the electron-momentum dependence of the EPC can be a powerful indicator of the microscopic structure of coupled electron-phonon systems and help distinguish the couplings of phonons to the kinetic energy and the potential of the electrons.

\noindent\includegraphics[width=\textwidth]{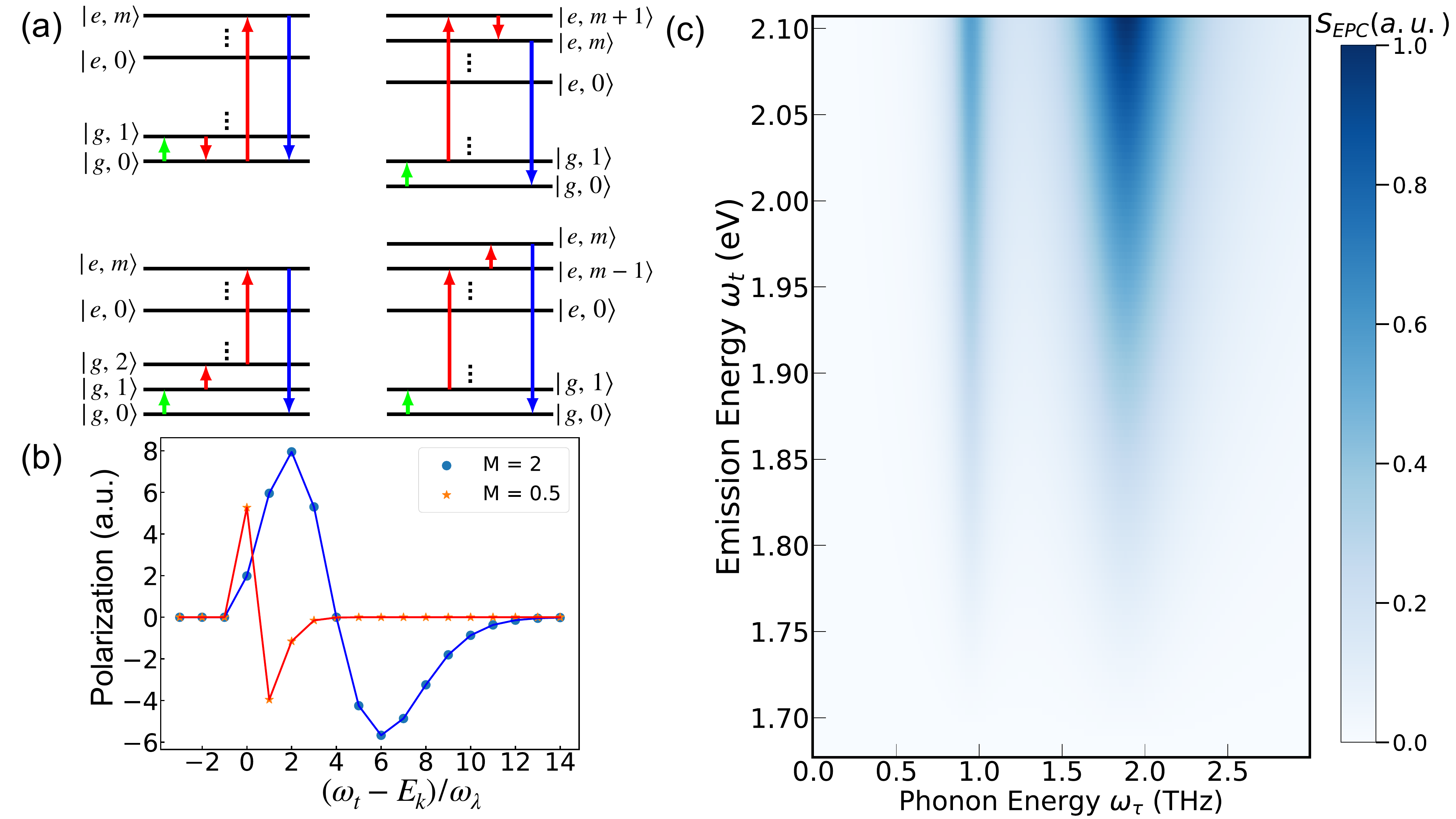}
\noindent {\bf Figure 3. Analytic Calculation of 2D EPC Spectrum} (a) Possible quantum pathways considering the non-rephasing contributions ($\omega_\tau >0, \omega_t>0$) of the third-order polarization  for the interacting electron-phonon system. The green arrow denotes the interaction with the THz pulse, the two red arrows denote the two-photon process caused by the NB and BB pulses, and the blue arrow denotes the emission, respectively. Interband transitions can involve multi-phonon excitations because of the EPC. (b) The resulting 2D EPC polarization contributed from all possible pathways in (a) for a given detection energy $\omega_t=\varepsilon_\mathbf{k}+m\omega_\lambda$. (c) The calculated 2D EPC spectrum, $S_{\rm EPC} (\omega_t,\omega_\tau)\sim P_{\rm EPC}(\omega_t,\omega_\tau)$, based on the total polarization given by Eq. (\ref{Ptot}).

\vspace{3cm}

Since we are able to directly measure and extract the phonon-mode- and electron-energy-resolved EPC strength at a fixed condition (e.g. temperature), we are now in a position to further explore how the EPC evolves across a phase transition.  MAPI is known to undergo a structural phase transition from a tetragonal (I4/mcm) to an orthorhombic (Pnma) phase below $T_c \sim$160 K \cite{Herz_2016} and to have optical and electronic properties that are highly sensitive to temperature changes \cite{Kong_2015,Singh_2016, Parrott_2016}. Due to the symmetry reduction in the low temperature phase, the degeneracy of both 1 and 2 THz modes in the high temperature phase is lifted and both modes split into two modes below $T_c$ \cite{La_o_vorakiat_2015}. Both splittings occur in such a way that additional modes with slightly lower frequencies (denoted by $\sim$ 0.7 THz and $\sim$1.5 THz in Fig. 4(a) appear besides the original modes (denoted by $\sim$1 and $\sim$2 THz in Fig. 4(a)). Note that all of the mode frequencies increase slightly with temperature. In 2D EPC, as shown in Fig. 4(a) (also 2D EPC spectra in Fig. S6), the EPC of each of these modes can be clearly distinguished. However, the original modes keep dominating the coupling strengths (with the 1 THz mode having a consistently stronger EPC than the 2 THz mode). Given the mode degeneracy above $T_c$, the combined contributions from the \textit{e.g.} $\sim$ 1.5 and $\sim$ 2 THz phonons to the EPC strength at $\sim$ 2 THz is overall reduced across the transition from orthorhombic (below $T_c$) to tetragonal (above $T_c$) phase.

Finally, another intriguing attribute of the phonon-mode- and electron-energy-specific EPC strength arises when we choose different polarization combinations of the three incoming pulses. Given the fixed polarization direction of the THz pulse (denoted by X), we can select the polarization directions of the NB, BB pulses and the signal field to be parallel (X) or perpendicular (Y) to that of the THz pulse (Fig. S11). As is evident from Fig. 4(b), the black (all parallel) and the red (cross polarization between the THz and the two optical pulses) curves reveal a contrast between coupling of the 1 THz mode and that of the 2 THz mode. With the cross polarization condition, the signal of the 1 THz mode is suppressed largely compared to that of the 2 THz mode, which highlights a clear anisotropic nature of EPC only for 1 THz case at $\sim$2 eV.  Further theoretical studies will be beneficial to gain microscopic insights into the geometrical nature of the phonon mode-specific EPC and  its sensitivity to the electronic momentum in terms of the relevant symmetries.

\vspace{1cm}
\noindent\includegraphics[width=\textwidth]{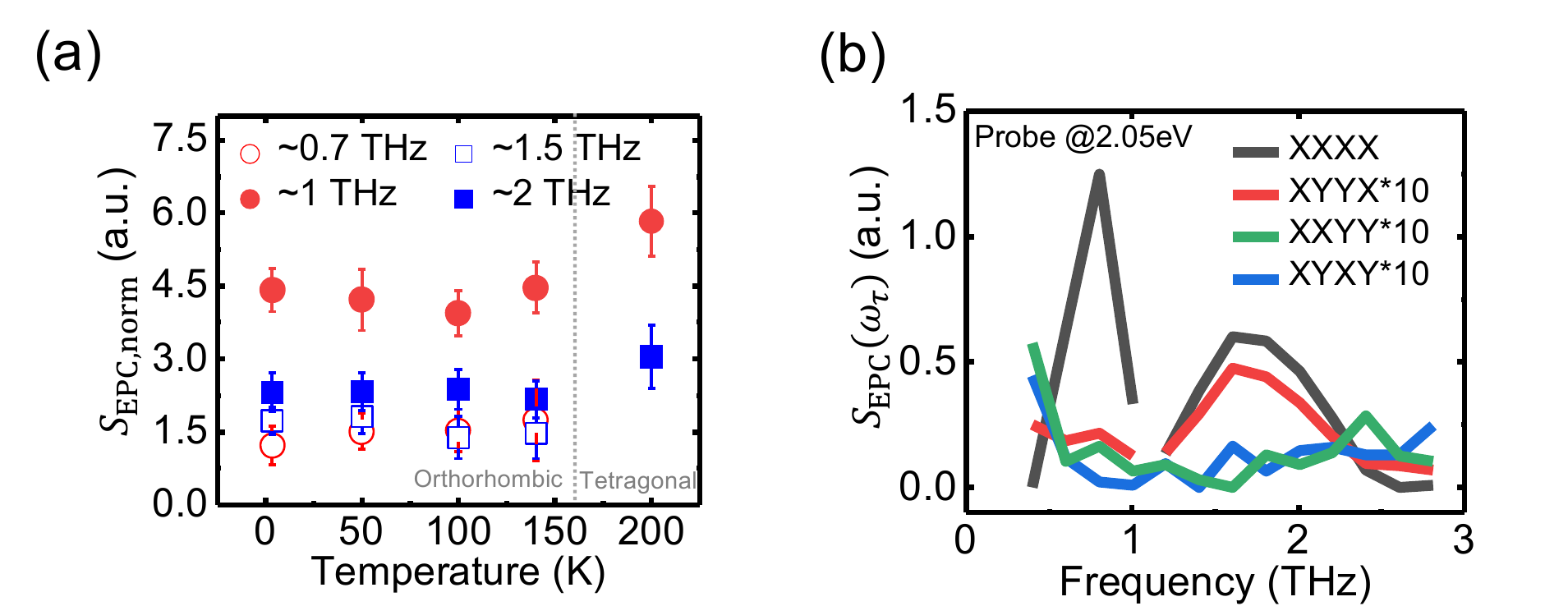}
\noindent {\bf Figure 4. Temperature and Polarization Dependence of Electron-Phonon Coupling} (a) The isolated contributions of phonon mode resolved EPC at the electron energy of 2.05 eV to the 2D EPC spectrum, $S_{\rm EPC, norm}$, of MAPI measured at temperatures of 3.7, 50, 140, 200 K. (b) Slice of the 2D EPC spectrum of MAPI, $S_{\rm EPC}(\omega_\tau)$, at electron energy 2.05 eV  measured by different polarization combinations of incident THz, NB, and BB pulses and signal fields.
\vspace{1cm}

\section{Conclusion}\label{sec4}

In conclusion, we have designed and demonstrated a novel experimental technique, namely 2D EPC spectroscopy, which allows for direct extraction of the phonon-mode-resolved EPC strength as a function of electron energy. We have shown that the measured 2D response function is a function of the EPC strength and that the signal vanishes in the absence of a coupling. In our demonstration of the technique on MAPI, among the two pronounced phonon modes at 1 and 2 THz, the lower frequency mode has an overall stronger and anisotropic EPC. 

Compared to established experimental techniques that are sensitive to EPC, 2D EPC spectroscopy has certain limitations and unique strengths. First of all,  2D EPC spectroscopy probes the differences in EPC strength for different electronic bands. This is tied to the fact that our technique directly probes the bare EPC matrix elements rather than indirect consequences of the coupling like the lifetime of electronic quasiparticles, which require an average over final electronic scattering states. Because 2D EPC spectroscopy directly probes correlated electron-phonon excitations, it remains unaffected by spurious effects such as electron-electron interactions, which may lead to ambiguity in all-electronic measurements. The band-resolved EPC strength therefore can be obtained by comparing 2D EPC with other experiments or ab-initio calculations. Moreover, the EPC difference is itself a relevant physical observable, for instance, it determines the energy renormalization and nonradiative lifetime of excitons due to phonons \cite{Marini2008}. Secondly, 2D EPC spectroscopy probes optically-active phonons at zero momentum. This may be particularly relevant for determining the EPC strength near instabilities driven by $\B q=0$ phonons, such as structural phase transitions, charge-density wave systems, or excitonic insulators (see, e.g., \cite{Chen2023}). Thirdly, our 2D EPC spectroscopy measures the EPC strength as a function of electron energy, and for a known band structure, we can therefore extract the electron-momentum dependence of the EPC strength  averaged over isoenergy surfaces.

Most importantly, the unique combination of electron-momentum ($\B k$) and phonon-mode resolution of our technique enables us to detect signatures of the microscopic origin of the EPC. The strong $\B k$-dependence along with the vanishing of the EPC strength at $\B k=0$ which we identify in MAPI is powerful evidence for a coupling of phonons to the kinetic energy rather than the potential. This makes our technique ideally suited for experimentally distinguishing nonlocal SSH-type couplings from local Holstein couplings. Incidentally, the contrast between both models is most pronounced at phonon momentum $\B q=0$, where our technique operates (see Supplementary Material). This is tied to the fact that at finite phonon momentum the interference of electron and lattice waves leads to a partial cancellation of the $\B k$-dependence.

Hence, 2D EPC spectroscopy yields complementary information compared to traditional techniques sensitive to EPC such as ARPES, heat capacity or transport measurements and can therefore play an important role in improving our understanding of EPC. Our work paves the way for studying the evolution of mode-specific EPC under changes of thermodynamic or mechanical conditions, as well as phonon-mediated light-induced ultrafast control of condensed materials. The combination of material specific calculation and this experimental approach is expected to provide deeper insight into the nature of EPCs.

\section*{Materials and Methods}

\subsection*{Experimental Details}
The output of an amplified femtosecond Ti:Sapphire laser (1 kHz repetition rate, 50fs pulse duration) is split into $\sim $ 1, 1, and 0.1 mJ for generating THz, NB, and BB NIR pulses, respectively. The broadband THz pulses are generated via femtosecond two-color laser focusing in long air-plasma filaments \cite{Bradler_2009}. The NB NIR pulses with bandwidth of $\sim $20 cm$^{-1}$ are obtained by sending the output of a commercial optical parametric amplifier into a homebuilt pulse shaper, which spatially selects spectral components by combination of a ruled reflective diffraction grating and a slit. The BB NIR pulses are produced via white light generation at 4 mm thick YAG (yttrium aluminum garnet) crystal \cite{You_2012} (Fig. S2). 

We combine the NB and BB pulses with a dichroic beam combiner in such a way that they collinearly propagate through the sample. The pulse energies of NB and BB pulses at the sample position are of $\sim$ 1 - 2 $\mu$J and $\sim$ 0.02 $\mu$J, respectively, and are focused by an off-axis parabolic mirror with focal length of 15 cm. The THz pulses are focused by another off-axis parabolic mirror with focal length of 10 cm. To measure the signal in a phase-resolved fashion, we adopt the heterodyne detection. The LO pulses are, therefore, generated after the sample position by sum frequency generation of NB and BB pulses at a 10 $\mu$m thick BBO ($\beta$-barium borate) crystal. Since we set NB and BB to overlap in time, when they pass through the sample and reach the BBO crystal, new pulses with a frequency of $\omega_{\rm NB}+\omega_{\rm BB}$ will be generated. The thus generated LO and signal fields are collimated via a CaF$_2$ lens and collinearly propagate through an analyzer polarizer. The intensity of LO interfered with and without the signal field are spatially separated by a vibrating motorized mirror (galvanometer), spectrally resolved, and imaged separately onto a CCD camera. The area of the setup around the THz beam pathway was enclosed in a box and purged with dry nitrogen. The sample is kept in a cryostat. 

\subsection*{Data Analysis}

The 2D EPC spectroscopy employs the third-order nonlinear optical process, and the generated 2D EPC signal field, $E_{\rm EPC}$, is proportional to the induced polarization to third order of the electric field of the incident lights: 
\begin{multline}\label{E_EPC}
E_{\rm EPC}(\mathbf{r}, t,\tau) \propto P_{\rm EPC}(\mathbf{r}, t,\tau) \\
\propto
\iiint d \mathbf{k}_1 d \mathbf{k}_2 d \mathbf{k}_3 \iiint d \omega_1 d \omega_2 d \omega_3 S_{\rm EPC}(\mathbf{k}_1,\mathbf{k}_2,\mathbf{k}_3, \omega_1 \pm \omega_2 \pm \omega_3, \omega_1 \pm \omega_2, \omega_1)  \\E_{\rm THz}(\mathbf{k}_1,\omega_1)e^{-i\omega_1 \tau+i \mathbf{k}_1 \cdot \mathbf{r}} E_{\rm BB}(\mathbf{k}_2,\omega_2)E_{\rm NB}(\mathbf{k}_3,\omega_3)e^{-i\omega_s t+i \mathbf{k}_s \cdot \mathbf{r}}
\end{multline}
where $t$ is the time variable of the signal field $E_{\rm EPC}$ and $\tau$ is a time delay of the two IR pulses relative to the THz pulse. The electric fields $E_{\rm THz}(\omega_1)$, $E_{\rm NB}(\omega_2)$ and $E_{\rm BB}(\omega_3)$ correspond to the THz, NB and BB pulses at the frequencies $\omega_1$, $\omega_2$ and $\omega_3$, respectively. Here, $\omega_s \equiv \omega_1 \pm \omega_2 \pm \omega_3$ and $\mathbf{k}_s \equiv \mathbf{k}_1 \pm \mathbf{k}_2 \pm \mathbf{k}_3$. The nonlinear response function $S_{\rm EPC}$ is expressed in time domain in terms of correlation functions with dipole moment operators $\mu$,
\begin{equation}\label{nonlinear S}
\begin{split}
S_{\rm EPC}\left(t_{3}, t_{2}, t_{1}\right)  =&  \theta\left(t_{1}\right) \theta\left(t_{2}\right) \theta\left(t_{3}\right) \\
&\times\left\langle\left[\left[\left[\mu\left(t_{3}+t_{2}+t_{1}\right), \mu\left(t_{2}+t_{1}\right)\right], \mu\left(t_{1}\right)\right], \mu(0)\right] \rho(-\infty)\right\rangle 
\end{split} 
\end{equation}
where $\theta\left(t\right)$ is Heaviside function and $\rho(-\infty)$ is the equilibrium reduced density matrix for the system eigenstates at $t_0$ (Fig. 1(a)) \cite{mukamel_1999}. The relevant diagrams for Eq.~(\ref{nonlinear S}) are shown in Fig. 1(c). Each term of the commutators indicates a distinctive Liouville space pathway at a given phase matching direction,  $\mathbf{k}_s$. Then, the raw measured spectrum in the frequency domain, $I_{\rm EPC}(\omega_t,\omega_\tau)$, contains convoluted contributions from each of the incoming fields themselves, the interference of signals from different phase matching directions, and the relative phase shift of the incoming fields at the interface, apart from the response function, $S_{\rm EPC} (\omega_t,\omega_\tau)$, that needs to be isolated.

Technically, the combined signal field and LO field are guided to the spectrometer (\textit{i.e.}, heterodyne detection), where they are dispersed by the grating. This dispersion is mathematically described by a Fourier transform of the field, $E_{\rm tot}(t,\tau)=E_{\rm EPC}(t, \tau)+E_{\rm LO}(t)$ over time $t$. The signal intensity measured by the square-law detector (CCD camera) for the detection frequency $\omega_t \geq 0$ is equivalent to $I_{\rm EPC}(\omega_t,\tau) \propto \int_T dt[E_{\rm tot}(\omega_t,\tau)e^{-i\omega_t t}+E_{\rm tot}(-\omega_t,\tau)e^{i\omega_t t}]^2$
with the accumulation time period of $T$. By squaring the term under the integral, neglecting terms oscillating at very high frequency $(\sim 2 \omega_t)$ and subtracting the homodyne terms of the form $E_{\rm EPC}(\omega_t,\tau)E_{\rm EPC}(-\omega_t,\tau)$ and $E_{\rm LO}(\omega_t,\tau)E_{\rm LO}(-\omega_t,\tau)$ we obtain the heterodyne-detected part of the measured signal (Fig. S3(a)). Through further Fourier transform over the time delay $\tau$ of the THz pulse relative to the NB and BB pulses, the measured (heterodyne-detected) 2D EPC spectrum is given by: 
\begin{equation}\label{S6}
\begin{split}
I_{\rm EPC}(\omega_t,\omega_\tau) 
& \propto FT[E_{\rm EPC}(\omega_t,\tau)]E_{\rm LO}(-\omega_t)+FT[E_{\rm EPC}(-\omega_t,\tau)]E_{\rm LO}(\omega_t) \\
&=E_{\rm EPC}(\omega_t,\omega_\tau)E_{\rm LO}(-\omega_t)+E_{\rm EPC}(-\omega_t,\omega_\tau)E_{\rm LO}(\omega_t)
\end{split}
\end{equation}

Meanwhile, from the Fourier transform of \eqref{E_EPC} along $t$, $E_{\rm EPC}(\omega_t,\omega_\tau)$ can be written:
\begin{equation}\label{S7}
\begin{split}
E_{\rm EPC}(\omega_t,\omega_\tau) &= FT[E_{\rm EPC}(\omega_t,\tau)]\\
& \propto \iiint d \omega_1 d \omega_2 d \omega_3 S_{\rm EPC}(\omega_1 + \omega_2 + \omega_3, \omega_1+\omega_2, \omega_1) \\  & E_{\rm THz}(\omega_1) \delta (\omega_\tau-\omega_1) E_{\rm BB}(\omega_2) E_{\rm NB}(\omega_3)\delta(\omega_t-\omega_s)
\end{split}
\end{equation}

Since we use a NB NIR pulse with the frequency $\omega_{\rm NB}$ in the experiment, we approximate the NB NIR spectrum by a delta function: $E_{\rm NB}(\omega_3)=\delta(\omega_{\rm NB}-\omega_3) + \delta(\omega_{\rm NB}+\omega_3)$. By performing the integration of \eqref{S7} and use the experiment condition that $E_{\rm BB}(\omega_t+\omega_{\rm NB}-\omega_\tau)=0$   ($\omega_\tau \ll \omega_t+\omega_{\rm NB}$), we obtain:
\begin{equation} \label{S9}
E_{\rm EPC}(\omega_t,\omega_\tau) \propto S_{\rm EPC}(\omega_t,\omega_t-\omega_{\rm NB},\omega_\tau) E_{\rm THz}(\omega_\tau)E_{\rm BB}(\omega_t-\omega_{\rm NB}-\omega_\tau) 
\end{equation}
Similarly, for $E_{\rm EPC}(-\omega_t,\omega_\tau)$ we obtain:
\begin{equation} \label{S10}
    E_{\rm EPC}(-\omega_t,\omega_\tau) \propto S_{\rm EPC}(-\omega_t,-\omega_t+\omega_{\rm NB},\omega_\tau) E_{\rm THz}(\omega_\tau)E_{\rm BB}(-\omega_t+\omega_{\rm NB}-\omega_\tau)
\end{equation}
By substituting \eqref{S10} and \eqref{S9} into \eqref{S6} we obtain for the 2D EPC spectrum:
\begin{multline} \label{2D EPC spectrum}
    I_{\rm EPC}(\omega_t,\omega_\tau) \propto S_{\rm EPC}(\omega_t,\omega_t-\omega_{\rm NB},\omega_\tau) E_{\rm THz}(\omega_\tau)E_{\rm BB}(\omega_t-\omega_{\rm NB}-\omega_\tau) E_{\rm LO}(-\omega_t) \\ + S_{\rm EPC}(-\omega_t,-\omega_t+\omega_{\rm NB},\omega_\tau) E_{\rm THz}(\omega_\tau)E_{\rm BB}(-\omega_t+\omega_{\rm NB}-\omega_\tau) E_{\rm LO}(\omega_t)
\end{multline}

Due to the collinear geometry of our experiment, the two terms in \eqref{2D EPC spectrum}, i.e. signals from different phase matching conditions, namely, non-rephasing pathways ($\omega_\tau>0, \omega_t >0$) and rephasing pathways ($\omega_\tau>0, \omega_t <0$) are spatially not separated and spectrally very close ($\omega_\tau \ll \omega_t$). This copropagation results in the interference of the two signals in the raw 2D spectrum (Fig. S3(b)). To separate these two types of contributions, we employ the fact that the two terms in \eqref{2D EPC spectrum} respond to the time delay $t'$ between LO and the signal field in an opposite way:
\begin{equation} \label{I_EPC_ks}
    I_{\rm EPC}^{\rm nr}(\omega_t,\omega_\tau) \propto S_{\rm EPC}(\omega_t,\omega_t-\omega_{\rm NB},\omega_\tau) E_{\rm THz}(\omega_\tau)E_{\rm BB}(\omega_t-\omega_{\rm NB}-\omega_\tau) E_{\rm LO}(-\omega_t)e^{i\omega_t t'} 
\end{equation}
\begin{equation}
    I_{\rm EPC}^{\rm r}(\omega_t,\omega_\tau) \propto  S_{\rm EPC}(-\omega_t,-\omega_t+\omega_{\rm NB},\omega_\tau) E_{\rm THz}(\omega_\tau)E_{\rm BB}(-\omega_t+\omega_{\rm NB}-\omega_\tau) E_{\rm LO}(\omega_t)e^{-i\omega_t t'}
\end{equation}
where $I_{\rm EPC}^{\rm nr}$ ($I_{\rm EPC}^{\rm r}$) indicates the signal intensity contributed from the non-rephasing (rephasing) pathway.
Hence, in the inverse Fourier transform along $\omega_t$ of the 2D spectrum, contributions from $I_{\rm EPC}^{\rm nr}$ and $I_{\rm EPC}^{\rm r}$ are temporally separated and one could select one of the two phase matching directions (Fig. 2(a)) \cite{10.1063/5.0047918}.

Once we isolate the 2D spectrum of non-rephasing contributions, we could now utilize from Eq. \eqref{I_EPC_ks} that the 2D response function $S_{\rm EPC}(\omega_t,\omega_\tau)$ can be obtained by dividing the measured 2D spectrum $I_{\rm EPC}^{\rm r}(\omega_t,\omega_\tau)$ directly by the incident field spectra, $E_{\rm THz}(\omega_\tau)$ and $E_{\rm LO}(\omega_t)$ (each spectrum given by Fig. S2(a) and S2(c)) . Note that we do not scan the time between NB and BB pulses, so the spectrum of BB pulse is not needed.

Finally, the relative phase differences of the incident fields at the sample are compensated by using the 2D EPC spectrum of a reference sample measured in the experimentally identical condition. A reference sample must have no resonance, so that the response function spectrum $S^{\rm ref}_{\rm EPC}(\omega_t,\omega_\tau)$ of the reference sample is a real constant over the spectral range (which $\omega_t$ and $\omega_\tau$ cover). Then, phase-shifted incident fields, i.e. $\tilde{E}_{\rm THz}(\omega_\tau) = E_{\rm THz}(\omega_\tau)\exp{i\omega_\tau t''}$ etc., where $t^{"}$ is relative phase difference, can be cancelled out by \cite{10.1063/1.3135147}:
\begin{equation}
    S^{\rm samp}_{\rm EPC}(\omega_t,\omega_\tau) \propto
    \frac{I^{\rm samp}_{\rm EPC}(\omega_t,\omega_\tau)}{S^{\rm ref}_{\rm EPC}(\omega_t,\omega_\tau) \tilde{E}_{\rm THz}(\omega_\tau) \tilde{E}_{\rm LO}(\omega_t)}
\end{equation}
In this experiment, we use 1 $\mu$m thick SiN membrane as the reference sample (Fig. S7$\sim$S9).

{\subsection*{Theoretical Model for 2d EPC spectroscopy}}

Here, we theoretically calculate the 2d spectrum for the generic electron-phonon model in Eq.~(2) of the main text. In the Supplementary Sec. 1.4 we calculate the induced polarization, $P_{\mathrm{EPC, tot}}(\omega_t)$, by considering the contributions from all possible diagrams and intermediate states.
We find that the contributions from diagrams $R_2^*$ and $R_3^*$ to the polarization cancel for our specific pulse sequence, and we only need to evaluate the contribution from diagram $R_4$. All possible pathways relevant for emission at frequency $\varepsilon_\B k+m\omega_{\lambda}$ of the non-rephasing condition contain the sequence of states $\ket{g,0}\to \ket{g,1}\to \ket{i}\to \ket{e_\B k,m}\to \ket{g,0}$ and we have to sum over all intermediate states $\ket{i}$ (see Fig. 3(a)). Since the transition dipole matrix element of conduction and valence band is nonzero, the state $\ket{i}$ must also be in the conduction  or valence band, as otherwise one of the transition matrix elements $\mu_{i\nu,g\nu'}$ or $\mu_{i\nu,e\nu'}$ would be zero by symmetry. The detailed evaluation all pathways is presented in Supplementary Sec. 1.5. Summing over all intermediate states, we obtain  $P_{\rm EPC,\mathbf{k}}(\omega_t=\varepsilon_\B k+m\omega_{\lambda})\propto e^{-M_{\B k,\lambda}^{2}} (M_{\B k,\lambda})^{2m}(M_{\B k,\lambda}^2-m)/m!$, 
for a given electron momentum $\B k$ and phonon number $m$ of the mode $\lambda$. The polarization at a detection frequency $\omega_t$ is then given by summing over all phonon numbers and all electronic momenta of the final state,
\begin{align}
P_{\rm EPC,{\rm tot}}(\omega_t,\omega_\tau)=&
\int d \varepsilon_\B k \nu(\varepsilon_\B k)
P_{\rm EPC,\mathbf{k}}(\omega_t,\omega_\tau)\\
\propto&\int d \varepsilon_\B k \nu(\varepsilon_\B k)
\left|\braket{e_\B k|{\mu}_{e}|g}\right|^{2} \ \left|\braket{\chi_{g,\lambda}^{0}|\mu_p|\chi_{g,\lambda}^1}\right|^2
 \frac{\Gamma_\lambda}{(\omega_\tau-\omega_\lambda)^2+\Gamma_\lambda^2}\notag\\
&\times \sum_m
 C_{m,\B k} e^{-M_{\B k,\lambda}^{2}} (M_{\B k,\lambda})^{2m}\frac{(M_{\B k,\lambda}^2-m)}{m!}\pi\delta(\varepsilon_\mathbf{k}+m \omega_{\lambda}-\omega_t).\label{ptot2}
\end{align}

Here $\nu(\varepsilon)$ is the electronic density of states,  $\Gamma_\lambda$ is the intrinsic broadening of the phonon mode $\lambda$, $\omega_\tau$ is the Fourier transform of the delay time $\tau$ between the first pulse and the NB and BB pulses, and $C_{m,\B k}$ is given by
\begin{align}
  C_{m,\B k}=
 &
\frac{1}{-\sigma^{2}_{3} [\omega_{3} - (\varepsilon_\mathbf{k} + m \omega_{\lambda})] + \sigma^{2}_{2}(\omega_{2} + \omega_{\lambda})}\notag\\
 &+\frac{1}{-\sigma^{2}_{3} [\omega_{3} - (\varepsilon_\mathbf{k} + (m -2) \omega_{\lambda})] + \sigma^{2}_{2}(\omega_{2} - \omega_{\lambda})}+(\sigma_2,\omega_2\leftrightarrow \sigma_3,\omega_3)
\end{align}
where $\omega_{2/3}$ and $\sigma_{2/3}$ are the central frequencies and time-domain widths of the two IR pulses, which are assumed to be Gaussian. Writing this expression we have assumed that the frequencies of the incoming pulses are chosen to be resonant with the respective transitions. We can perform the sum over $m$ analytically, if we ignore the weak dependence  of $C_{m,\B k}$ on $m$,  \textit{i.e.,} $C_{m,\B k}\simeq C_{0,\B k}$, and we obtain
\begin{align}
 e^{-M_{\B k,\lambda}^{2}} \sum_{m=0}^\infty&
 (M_{\B k,\lambda})^{2m}\frac{(M_{\B k,\lambda}^2-m)}{m!}\delta(\varepsilon_\mathbf{k}+m \omega_{\lambda}-\omega_t)\notag\\
 =&
e^{-M_{\B k,\lambda}^{2}} \sum_{m=0}^\infty (M_{\B k,\lambda})^{2m}\frac{M_{\B k,\lambda}^2}{m!}[\delta(\varepsilon_\mathbf{k}+m \omega_{\lambda}-\omega_t)-
 \delta(\varepsilon_\mathbf{k}+(m+1) \omega_{\lambda}-\omega_t)]\\
 \simeq&
 -\omega_\lambda \delta'(\varepsilon_\mathbf{k}-\omega_t) e^{-M_{\B k,\lambda}^{2}} \sum_{m=0}^\infty (M_{\B k,\lambda})^{2m}\frac{M_{\B k,\lambda}^2}{m!}=
 -\omega_\lambda \delta'(\varepsilon_\mathbf{k}-\omega_t) M_{\B k,\lambda}^2,
\end{align}
where $\delta'(\omega)\simeq [\delta(\omega+\omega_\lambda)-\delta(\omega)]/\omega_\lambda$ denotes the derivative of a $\delta$-function.
Substituting this into Eq.~(\ref{ptot2}), we find
\begin{align}
P_{\rm EPC,{\rm tot}}(\omega_t,\omega_\tau)
\propto&\int d \varepsilon_\B k \nu(\varepsilon_\B k)
\left|\braket{e_\B k|{\mu}_{e}|g}\right|^{2} \ \left|\braket{\chi_{g,\lambda}^{0}|\mu_p|\chi_{g,\lambda}^1}\right|^2
 \frac{\Gamma}{(\omega_\tau-\omega_\lambda)^2+\Gamma^2}\notag\\
&\times C_{0,\B k} \pi
 \omega_\lambda \delta'(\varepsilon_\mathbf{k}-\omega_t) M_{\B k,\lambda}^2
\end{align}

In this expression, only $ \nu(\varepsilon_\B k)$ and $M_{\B k,\lambda}^2$ have a strong dependence on $\B k$, and, thus, we can further simplify the result as Eq.~(\ref{analytics}) of the main text with the definition
\begin{align}
 C(\varepsilon_\B k)\propto\left|\braket{e_0|{\mu}_{e}|g}\right|^{2} \ \left|\braket{\chi_{g,\lambda}^{0}|\mu_p|\chi_{g,\lambda}^1}\right|^2
  C_{0,\B k}.
\end{align}
In order to determine the EPC strength from the  experimental data, we fit the normalized 2D EPC spectrum for a given phonon mode, $S_{\mathrm{EPC,norm}}^{\lambda} (\omega_t)$, in Fig.~2(d). In our theoretical model, the normalized spectrum, $S_{\mathrm{EPC,norm}}^{\lambda} (\omega_t)$, corresponds to the intensity $I_{\rm EPC}\propto \omega_t P_{\rm EPC, tot}$ divided by the THz absorbance $\propto \omega_\lambda\left|\braket{\chi_{g,\lambda}^{0}|\mu_p|\chi_{g,\lambda}^1}\right|^2
 \Gamma /[(\omega_\tau-\omega_\lambda)^2+\Gamma^2]$ and the IR absorbance $\propto \omega_t \nu(\omega_t) \left|\braket{e_0|{\mu}_{e}|g}\right|^{2} $ \cite{mukamel_1999}. We then use Eq.~(\ref{invert}) to determine the EPC strengths. We perform the integration by assuming the density of states to be a power law $\nu(\omega_t)\propto (\omega_t-\epsilon_\Gamma)^\alpha$  and by fitting the experimental data for $S_{\mathrm{EPC,norm}}^{\lambda} (\omega_t)$ in Fig.~2(d) with a polynomial. Specifically, we fit the  spectrum of the 1 THz phonon mode  by a quadratic function and that of the 2 THz mode by a linear function,  assuming both functions cross zero at the gap edge, which we find to be $\varepsilon_{\Gamma} = 1.68\,$eV. This allows us to obtain $M_{\varepsilon_{\B k},\lambda}$ which is shown in Fig.~2(e). We verify that the coupling strength obtained from the analytical expression (\ref{invert}) indeed reproduces the experimental data by plotting the full numerical result based on Eqs.~(\ref{Ptot}) and (\ref{ptot2}) in Fig.~2(d). The broadening of the phonon modes are obtained by fitting the THz absorbance in Fig.~2(c) with a Lorentz oscillator model of absorbance \cite{La_o_vorakiat_2015}, from which we find $\Gamma_{1} = 0.11$\,THz and $\Gamma_{2} = 0.27$\,THz. The oscillator strength of the 2 THz mode is found to be approximately four times higher than that of the 1 THz mode.

\clearpage

\backmatter

\bmhead{Supplementary information}

Materials and Methods\\
Supplementary Text\\
Figs. S1 to S12\\
References \textit{(1-38)}\\

\bmhead{Acknowledgments}

V.K.S. and F.P. acknowledge funding by the Deutsche Forschungsgemeinschaft (DFG) through the Collaborative Research Center TRR 288 (project number 422213477, project B09). H.K. and S.Q. thank the DFG for financial support through the Collaborative Research Center TRR 288 (project number 422213477, project B07). This work was also supported by the National Research Foundation (NRF) of Korea (grant number 2021R1A6A1A10042944). JJG gratefully acknowledges funding from the Alexander von Humboldt foundation.

\section*{Declarations}
\begin{itemize}
\item Ethics approval
: The authors declare no competing interests.

\item Availability of data and materials
: The data that support the findings of this study are available at Figshare 

\item Code availability
: The code that supports the findings of this study is available at Figshare 

\end{itemize}
\bibliography{References_2DEPC}
\clearpage


\includepdf{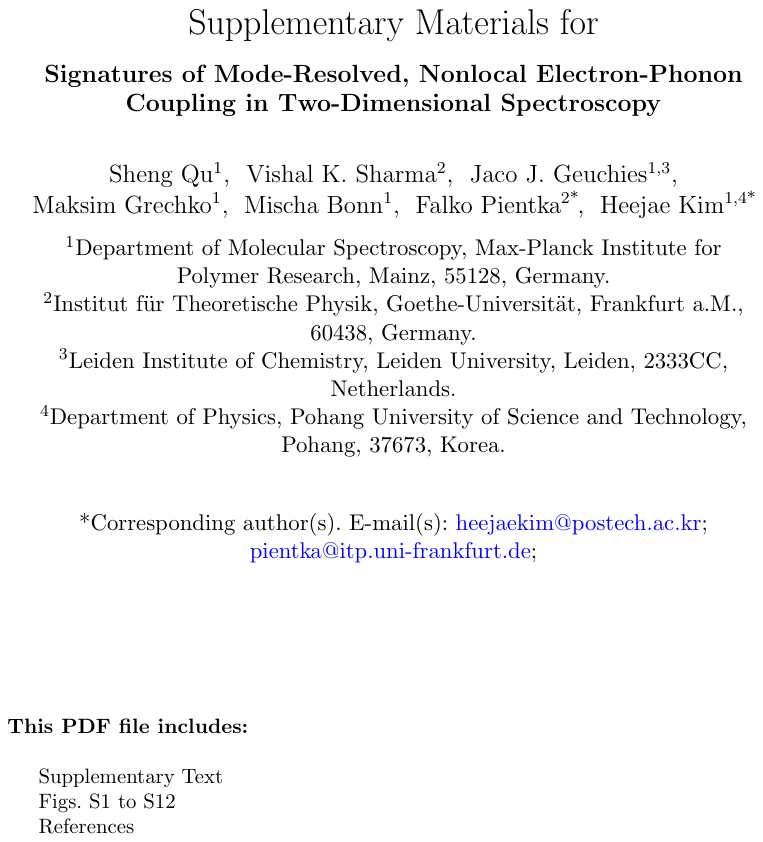}
\thispagestyle{empty}
\newpage
\renewcommand{\thepage}{A-\arabic{page}}

\setcounter{section}{0}
\setcounter{page}{1}
\def\theequation{A\arabic{equation}}
\setcounter{equation}{0}
\section{Supplementary Texts}\label{sec1}
\subsection{Perovskite film preparation}

Methylammonium iodide (MAI) was purchased from Greatcell Energy. Lead acetate (Pb(Ac)$_2$) 98\% purity was purchased from Tokyo Chemical Industry. Anhydrous dimethylformamide ($>$ 99.8\% purity) was purchased from Sigma Aldrich. Substrates (0.3mm-thick glass) were cleaned by sonication in water with soap for 30 minutes, afterwards rinsed with deionized water three times, and again sonicated in ethanol and acetone (both for 30 minutes). Afterwards, the substrates were dried with a nitrogen flow and subjected to UV-Ozone treatment (FHR UVOH 150 LAB, 250 W) for 20 minutes with an oxygen feeding rate of 1 L/min right before spincoating. Film preparation was carried out in a nitrogen purged glovebox. The precursor solution consists out of 477 mg (3 mmol) MAI and 325.3 mg of Pb(Ac)$_2$ (1 mol) in 1 mL DMF. The film preparation was done by depositing 50 $\mu$L of the precursor solution onto a cleaned fused silica substrate and spincoating it at 4000 rpm (ramp ± 1000 rpm/s) for one minute. Afterwards the film was left to dry at room temperature for 10 minutes and subsequently annealed on a 100$^\circ$C hotplate 
\textcolor{blue}{$[1,2]$}. The films were stored in the glovebox prior to their use.

\subsection{Separation of Rephasing and Non-rephasing 2D EPC spectra}
The non-rephasing and rephasing signals expressed in Eq. (18-19)  are separated as follows. As shown in Fig. S4.A, the NB NIR, BB NIR and the emitted Signal from the sample pass through a glass substrate to reach the BBO2 crystal. Due to the lower refractive index of NIR compared to Signal in glass, NB NIR and BB NIR arrive at the BBO2 crystal earlier than the Signal. Therefore, the LO generated at BBO2 by the sum frequency of NB and BB leads the Signal by a time difference of $t’$. The rephrasing and non-rephasing signals are then separated in time by $2t’$ as shown in Fig. S4.B. Then, by performing the inverse Fourier transformation along the electron energy ($\omega_t$) axis of the I($\omega_t$, $\omega_\tau$) spectrum shown in Fig. S3.B, we can obtain the rephrasing/non-rephasing spectra separated in time, as shown in Fig. S4.C. Next, we perform the FT back along the $\Delta$$t$-axis for the signals in the first and fourth quadrants individually, so that we can obtain the absolute spectra contributed from the non-rephasing and rephasing pathways separately, as shown in Fig. S5. Finally, by performing the phase correction procedure using silicon nitride (SiN$_x$) membrane (described below) for each separated spectrum, we further separate the real and imaginary spectra corresponding to rephasing and non-rephasing pathways (Fig. S6).

\subsection{Phase Correction of 2D EPC Spectra}

In Fig. S7., one could see a linear phase dispersion in the measured 2D EPC signal, which affects the line shape. To get rid of the instrument phase artifact, we implement a similar phase correcting method to that for the phase-resolved sum frequency generation spectroscopy 
\textcolor{blue}{$[3]$}. We use a 1$\mu$m-thick SiN$_x$ membrane (QX151000F, Norcada) which does not have any resonances so that one could use it as a reference sample to extract the pure instrument phase information. In this non-resonant SiN$_x$ measurement, we not only keep the beam configurations exactly the same compared with MAPI measurement, also insert a 0.3mm-thick glass substrate between SiN$_x$ membrane and 10 $\mu$m BBO crystal to keep the same time shift between signal and LO, which is critical for this accurate phase measurement. Here we assume that the difference of time delay between 2D EPC signal and LO induced by SiN$_x$ membrane (1$\mu$m-thick) and MAPI thin film ($\sim$300 nm-thick) is negligible. By utilizing the same analytical method mentioned in the main text, the time domain and first-quadrant frequency domain spectra of SiN$_x$ membrane are shown in  Fig. S8. \\
Consequently, the real and imaginary 2D EPC spectra of MAPI can be obtained by dividing the 2D spectrum of MAPI in a given phase matching direction (Fig. 2(c)) by the SiN$_x$ membrane reference spectra (Fig. S8). The phased EPC spectra I$_p$$'$($\omega_t$, $\omega_\tau$) is given by the equation:
\begin{equation}
{I'_p(\omega_t, \omega_\tau)\propto\frac{I' (\omega_t, \omega_\tau; MAPI)}{I' (\omega_t, \omega_\tau; SiN_x \; membrance)}}.
\label{eq1}
\end{equation}
The phase corrected real and imaginary part of 2D EPC spectra of MAPI contributed from non-rephasing pathways are shown in Fig. S9.

\subsection{Calculation of the polarization diagrams}

Our calculation of the polarization follows the formalism described in \textcolor{blue}{${[4]}$}. 
 The third-order polarization is given by the expression
\begin{align} 
    P(t, \mathbf{r}) = & \int_{0}^{\infty} dt_{3} \int_{0}^{\infty} dt_{2} \int_{0}^{\infty} dt_{1}  \mathbf{E}(t-t_{3})\mathbf{E}(t-t_{3} -t_{2}) \notag\\
    &\times\mathbf{E}(t-t_{3}-t_{2}-t_{1}) S^{(3)}(t_{3}, t_{2}, t_{1})\label{polarization}
\end{align}
where 
\begin{equation}\label{pathways}
\begin{split}
S^{(3)}\left(t_{3}, t_{2}, t_{1}\right) & =  \left(\frac{i}{\hbar}\right)^{3}\theta\left(t_{1}\right) \theta\left(t_{2}\right) \theta\left(t_{3}\right) 
\left\langle\left[\left[\left[\mu\left(t_{3}+t_{2}+t_{1}\right), \mu\left(t_{2}+t_{1}\right)\right], \mu\left(t_{1}\right)\right], \mu(0)\right] \rho(-\infty)\right\rangle \\& = \left(\frac{i}{\hbar}\right)^{3} \theta\left(t_{1}\right) \theta\left(t_{2}\right) \theta\left(t_{3}\right) \sum_{\alpha = 1}^{4} \left[ R_{\alpha} - R^{*}_{\alpha} \right]
\end{split} 
\end{equation}
Here the shorthand notation $R_{i}$ denotes different pathways corresponding to different terms in the nested commutator of the response function 
\begin{equation}
\begin{array}{l}
R_{1}\left(t_{3}, t_{2}, t_{1}\right)=\langle \mu\left(t_{1}\right) \mu\left(t_{1}+t_{2}\right) \mu\left(t_{1}+t_{2}+t_{3}\right) \mu(0) \rho \left(-\infty_{1}\right)\rangle, \\
R_{2}\left(t_{3}, t_{2}, t_{1}\right)=\langle \mu(0) \mu\left(t_{1}+t_{2}\right) \mu(t_{1}+t_{2}+t_{3}\rangle \mu\left(t_{1}\right) \rho(-\infty)\rangle, \\
R_{3}\left(t_{3}, t_{2}, t_{1}\right)=\left\langle \mu(0) \mu\left(t_{1}\right) \mu\left(t_{1}+t_{2}+t_{3}\right) \mu\left(t_{1}+t_{2}\right) \rho\left(-\infty\right)\right\rangle, \\
R_{4}\left(t_{3}, t_{2}, t_{1}\right)=\langle \mu\left(t_{1}+t_{2}+t_{3}\right) \mu\left(t_{1}+t_{2}\right) \mu\left(t_{1}\right) \mu(0) \rho(-\infty)\rangle
\end{array}
\end{equation}
A diagrammatic representation of all four terms is shown in Fig.~S12. As can be readily understood we only have to calculate the terms $R^*_{2,3}$ and $R_4$. The electric field is given by 
 \begin{align}
  \mathbf{E}(t) =\sum_{i=1}^3 \mathcal{E}_{i}(t-T_i) e^{-i\omega_{i} t} 
\end{align}
where $T_i$ is the time at which each pulse arrives at the sample. Note that for our experiment and using the notation of Fig.~1(a) of the main text, we have $T_1=-\tau$, $T_2=T_3=0$. The functions ${\cal E}_i(t)={\cal E}_i(0) \exp(-t^2/2\sigma_i^2)$ are the Gaussian envelopes of the pulses centered around zero. We only consider contributions where one photon is absorbed from all three pulses and we now have to associate the different pulses $\mathcal{E}_i$ with the three electric field factors in Eq.~(\ref{polarization}). As the first pulse $\mathcal{E}_1$ does not overlap with the other two, it has to be earliest of the three pulses (i.e. the term $\mathbf{E}(t-t_{3}-t_{2}-t_{1})$). The second and third pulse overlap and we therefore obtain two contributions for each diagram. For diagram $R_4$  we have two contributions $\B P_4$ and $\B P_4'$, where either the narrow band or the broad band pulse comes first. We find
\begin{align} 
    \mathbf{P}_{4}(t) = &\left(\frac{i}{\hbar}\right)^{3}
    \sum_{a,b,c,d} 
m_{abcd}  e^{i(-\omega_{1}-\omega_{2}-\omega_{3})t} e^{i\omega_{3}T_{3}}
    e^{i\omega_{2}T_{2}}
    e^{i\omega_{1}T_{1}} \notag\\
    &\times\int_{0}^{\infty} dt_{
    3}   \mathcal{E}_{3}(t-t_{3} -T_{3}) e^{-i(-\omega_{1}-\omega_{2}-\omega_{3}+\omega_{da})t_{3}}\notag\\
    & \times\int_{0}^{\infty} dt_{2}    \mathcal{E}_{2}(t-t_{3} - t_{2}-T_{2}) e^{-i(-\omega_{1}-\omega_{2}+\omega_{ca})t_{2} }\notag\\
    &\times
   \int_{0}^{\infty} dt_{1}  \mathcal{E}_1(t-t_{3}-t_{2}-t_{1}-T_{1}) e^{-i(-\omega_{1}+\omega_{ba})t_{1}} \label{P4}
\end{align}
where $  m_{abcd} =  \mu_{dc}\mu_{cb}\mu_{ba}\mu_{ad} $. The second contribution $\B P_4'(t)$ is obtained from the above expression by exchanging ${\cal E}_2,\omega_2\leftrightarrow {\cal E}_3,\omega_3$.

The integrands are vanishingly small except when the arguments of the envelopes ${\cal E}_i$ are close to zero, which happens when the integration times equal the delay times between the four pulses (the three probe pulses and the local oscillator), i.e.,  $t_3\simeq t-T_3$, $t_2\simeq T_3-T_2$, and $t_1\simeq T_2-T_1$. The delay times between the first and second pulse, $T_2-T_1$, and between the third and local oscillator pulse, $t-T_3$, are much longer than the temporal widths of the envelopes, $\sigma_i$, and we can therefore safely extend the domain of integration to the entire real axis for the $t_{1}$ and $t_3$ integration. The second and third pulse, however, are overlapping in time, $T_2\simeq T_3$, and the $t_2$ integral, therefore, cannot be extended to the entire real axis.
We proceed by solving the $t_1$ integration and we obtain
\clearpage

\begin{align} 
    \mathbf{P}_{4}(t) = &\sqrt{2\pi}\left(\frac{i}{\hbar}\right)^{3}\sigma_1 \mathcal{E}_{1}(0)
    \sum_{a,b,c,d} 
   m_{abcd}    e^{i(-\omega_{ba}-\omega_{2}-\omega_{3})t} e^{i\omega_{3}T_{3}}
    e^{i\omega_{2}T_{2}}
    e^{i\omega_{ba}T_{1}}
     e^{-\sigma_1^2(\omega_{1}-\omega_{ba})^2/2}\notag\\
    & \times \int_{0}^{\infty} dt_{2}   e^{-i(-\omega_{ba}-\omega_{2}+\omega_{ca})t_{2} }\notag\\
    &\times
         \int_{-\infty}^{\infty} dt_{
    3}    \mathcal{E}_{2}(t-t_{3} - t_{2}-T_{2}) \mathcal{E}_{3}(t-t_{3} -T_{3}) e^{-i(-\omega_{ba}-\omega_{2}-\omega_{3}+\omega_{da})t_{3}}
\end{align}
We now solve the Gaussian $t_3$ integral and setting $T_2=T_3$ we find
\begin{align} 
    \mathbf{P}_{4}(t) = &2\pi\left(\frac{i}{\hbar}\right)^{3}\frac{\sigma_1\sigma_2\sigma_3}{\sqrt{\sigma_2^2+\sigma_3^2}} \mathcal{E}_{1}(0) \mathcal{E}_{2}(0) \mathcal{E}_{3}(0)
    \sum_{a,b,c,d} 
   m_{abcd}  e^{-i\omega_{da}t}   e^{i(-\omega_{ba}+\omega_{da})T_2} e^{i\omega_{ba}T_{1}}
     \notag\\
    & \times     e^{-\sigma_1^2(\omega_{1}-\omega_{ba})^2/2}   e^{-(\omega_{2}+\omega_{3}-\omega_{db})^2\sigma_2^2\sigma_3^2/2(\sigma_2^2+\sigma_3^2)}\notag\\
    &\times
    \int_{0}^{\infty} dt_{2}   
   e^{i[(\omega_{bc}+\omega_{2})\sigma_2^2 -(\omega_{3}+\omega_{cd})\sigma_3^2]t_2 /(\sigma_2^2 + \sigma_3^2)}
 e^{   -t_2 ^2/2 (\sigma_2^2 + \sigma_3^2)}
\end{align}
where we have used 
\begin{align}
 e^{-i(-\omega_{ba}-\omega_{2}+\omega_{ca})t_{2} }&
    e^{-i(-\omega_{ba}-\omega_{2}-\omega_{3}+\omega_{da})(-\sigma_3^2 t_2/(\sigma_2^2 + \sigma_3^2))}\notag\\
    &\qquad
    =e^{i[(\omega_{bc}+\omega_{2})\sigma_2^2 -(\omega_{3}+\omega_{cd})\sigma_3^2]t_2 /(\sigma_2^2 + \sigma_3^2)}
\end{align}
The remaining integral has the form $\int_0^\infty dx \exp(-x^2/4u^2+i vx)=\sqrt{\pi}u\exp(-u^2v^2)(1+i {\rm erfi}(uv))$
where ${\rm erfi}(uv)$ denotes the imaginary error function. In this case we have $uv\sim [(\omega_{bc}+\omega_{2})\sigma_2^2 -(\omega_{3}+\omega_{cd})\sigma_3^2]/\sqrt{\sigma_2^2 + \sigma_3^2}\gg 1$ since we have $\sigma_2\gg \sigma_3,1/\omega_2$ for the pulse duration $\sigma_2$ of the narrow band pulse. Note that this condition is violated if the square brackets happens to cancel, i.e. if $\omega_2\sim \omega_{cb}$ and $\omega_3\sim \omega_{dc}$. This occurs if the second and third pulse are both resonant with a transition in our system, which we delibarately avoid by choosing the frequency of the narrow band pulse to be below the band gap. 
Using $uv\gg 1$ we can approximate the error function by its asymptote
\begin{align}
 \int_0^\infty dx \exp(-x^2/4u^2+i vx)\simeq2\sqrt{\pi}u\exp(-u^2v^2)+\frac{i }{v}\simeq\frac{i }{v}
\end{align}
In summary, we arrive at
\begin{align} 
    \mathbf{P}_{4}(t) \sim &
    \sum_{a,b,c,d} 
   m_{abcd}    e^{-i\omega_{da}t}   e^{i(-\omega_{ba}+\omega_{da})T_2} e^{i\omega_{ba}T_{1}}
     e^{-\sigma_1^2(\omega_{1}-\omega_{ba})^2/2}
     \notag\\
    & \times   e^{-(\omega_{2}+\omega_{3}-\omega_{db})^2\sigma_2^2\sigma_3^2/2(\sigma_2^2+\sigma_3^2)}
\frac{1}{(\omega_{bc}+\omega_{2})\sigma_2^2 -(\omega_{cd}+\omega_{3})\sigma_3^2}
\end{align}
We can readily obtain the contribution from $R^*_2$ by substituting $\omega_2\to -\omega_2$, $\omega_{ca}\to\omega_{bd}$, and $\omega_{da}\to\omega_{cd}$ in Eq.~(\ref{P4}) and using similar manipulation we find
\begin{align}
    \mathbf{P}_{2}(t)  \sim &
    \sum_{a,b,c,d}
     m_{abcd} e^{-i\omega_{cd}t}   e^{i(-\omega_{ba}+\omega_{cd})T_2} e^{i\omega_{ba}T_{1}}
     e^{-\sigma_1^2(\omega_{1}-\omega_{ba})^2/2}
     \notag\\
    & \times   e^{-(-\omega_{2}+\omega_{3}-\omega_{cb}+\omega_{da})^2\sigma_2^2\sigma_3^2/2(\sigma_2^2+\sigma_3^2)}
\frac{1}{(\omega_{da}-\omega_{2})\sigma_2^2 -(\omega_{bc}+\omega_{3})\sigma_3^2}
\end{align}
The contribution from $R^*_3$ follows by substituting $\omega_3\to -\omega_3$ and $\omega_{da}\to\omega_{cd}$ in Eq.~(\ref{P4}),
\begin{align}
    \mathbf{P}_{3}(t)  \sim &
    \sum_{a,b,c,d}
     m_{abcd}   e^{-i\omega_{cd}t}   e^{i(-\omega_{ba}+\omega_{cd})T_2} e^{i\omega_{ba}T_{1}}
     e^{-\sigma_1^2(\omega_{1}-\omega_{ba})^2/2}
     \notag\\
    & \times   e^{-(\omega_{2}-\omega_{3}-\omega_{cb}+\omega_{da})^2\sigma_2^2\sigma_3^2/2(\sigma_2^2+\sigma_3^2)}
\frac{1}{(\omega_{bc}+\omega_{2})\sigma_2^2 -(\omega_{da}-\omega_{3})\sigma_3^2}
\end{align}
We also have to add the terms $\B P'_{2,3,4}(t)$ which are obtained from the above expression by replacing $\sigma_2,\omega_2\leftrightarrow \sigma_3,\omega_3$. We see that $\B P_2(t)=-\B P'_3(t)$ and $\B P'_2(t)=-\B P_3(t)$ so the $R_2^*$ and $R_3^*$ diagrams cancel each other and only the $\B P_4$ and $\B P_4'$ contributions from $R_4$ remain.

\subsection{Summation over pathways in Fig. 3(a)}

We now sum the polarization $P_4+P_4'$ over all pathways shown in Fig.~3(a).
For all four pathways we have $\ket{a}=\ket{g,0}$, $\ket{b}=\ket{g,1}$, and $\ket{d}=\ket{e,m}$ and we only need to consider four possible intermediate states $\ket{c}=\{\ket{g,0},\ket{g,2},\ket{e,m-1},\ket{e,m+1}\}$.
The state $\ket{c}$ only enters the expression for $P_4+P_4'$ through the term
\begin{align}
m_{abcd}\Bigl[\frac{1}{(\omega_{bc}+\omega_{2})\sigma_2^2 -(\omega_{cd}+\omega_{3})\sigma_3^2}
+(\sigma_2,\omega_2\leftrightarrow \sigma_3,\omega_3)\Bigr]
\end{align}
Considering the two pathways in the top row of Fig.~3(a), i.e., $\ket{c_1}=\ket{g,0}$ and $\ket{c_2}=\ket{e,m+1}$, we see that the transition frequencies only differ by the substitution $\omega_{bc}\leftrightarrow \omega_{cd}$ and we conclude that the square bracket above has equal magnitude but opposite signs for the two pathways. We can readily see that the same is true for $\ket{c_3}=\ket{g,2}$ and $\ket{c_4}=\ket{e,m-1}$ and we thus obtain
\begin{align}
P_4(t,\tau)+P_4'(t,\tau)\sim &\sum_d f_{abd}(t,\tau)\Bigl[\frac{m_{abc_1d}-m_{abc_2d}}{(\omega_{bc_1}+\omega_{2})\sigma_2^2 -(\omega_{c_1d}+\omega_{3})\sigma_3^2}\notag\\
&+\frac{m_{abc_3d}-m_{abc_4d}}{(\omega_{bc_3}+\omega_{2})\sigma_2^2 -(\omega_{c_3d}+\omega_{3})\sigma_3^2}
+(\sigma_2,\omega_2\leftrightarrow \sigma_3,\omega_3)\Bigr]\label{p4minusp4p}
\end{align}
where
\begin{align}
 f_{abd}(t,\tau)= e^{-i\omega_{da}t}   e^{-i\omega_{ba}\tau}
     e^{-\sigma_1^2(\omega_{1}-\omega_{ba})^2/2}
     e^{-(\omega_{2}+\omega_{3}-\omega_{db})^2\sigma_2^2\sigma_3^2/2(\sigma_2^2+\sigma_3^2)}
\end{align}
We now evaluate the product of dipole matrix elements $m_{abcd}=\mu_{ba}\mu_{ad}\mu_{cb}\mu_{dc}$. The intraband phonon transition matrix element varies with phonon number as $ \braket{n,m|\mu|n,m-1}\propto \sqrt{m}$. The interband electronic transitions introduce the Franck-Condon factors through the phonon wavefunction overlaps. In our model, the $q=0$ phonon wavefunctions of the electronic excited states are simply the coherent states of the electronic ground state phonon operators (see Eqs. (4-5) of the main text). This means we can express the excited state phonons in terms of the ground state phonons using the displacement operator
\begin{align}
  \ket{\chi^m_{e,\lambda}}
=  e^{-(b_\lambda^\dag -b_\lambda)M}\ket{\chi^m_{g,\lambda}},
\end{align}
where we have dropped the indices of $M=M_{\B k,\lambda}=M_{\mathbf{k}, e,\lambda}-M_{\mathbf{k},g, \lambda}$, we assume $M$ to be real for simplicity,
and we have used $e^{-(b_\lambda^\dag -b_\lambda)M}b_\lambda^\dag =(b_\lambda^\dag +M)e^{-(b_\lambda^\dag -b_\lambda)M}$. The matrix elements of the displacement operator are given by \textcolor{blue}{[5]}

\begin{align}
 \braket{\chi^{m'}_{g,\lambda}|\chi^m_{e,\lambda}}=\sqrt{\frac{m'!}{m!}}M^{m-m'}e^{-M^2/2}L_{m'}^{(m-m')}(M^2),
\end{align}
where  $L^k_n(x)$ is the associated Laguerre polynomial. \\

Considering the transitions in Fig.~3(a), we specifically need the overlaps
\begin{align}
  \braket{\chi^{m}_{e,\lambda}|\chi^0_{g,\lambda}}
&  =\sqrt{\frac{1}{m!}}M^{m}e^{-M^2/2}
\\
\braket{\chi^{m+1}_{e,\lambda}|\chi^1_{g,\lambda}}
&=\sqrt{\frac{1}{(m+1)!}}M^{m}(-M^2+m+1)e^{-M^2/2}
\\
\braket{\chi^{m}_{e,\lambda}|\chi^2_{g,\lambda}}
&=\sqrt{\frac{1}{2m!}}M^{m-2}(M^4-2M^2m+(m-1)m)e^{-M^2/2}
\\
\braket{\chi^{m-1}_{e,\lambda}|\chi^1_{g,\lambda}}
&=\sqrt{\frac{1}{(m-1)!}}M^{m-2}(-M^2+m-1)e^{-M^2/2}
\end{align}
We can now evaluate the product of all four dipole matrix elements (involving two intraband and two interband transistion) for each of the four pathways
\begin{align}
 m_{abc_1d}&=A \frac{1}{m!}M^{2m}e^{-M^2}
\\
m_{abc_2d}&=A \frac{1}{m!}M^{2m}(-M^2+m+1)e^{-M^2}
\\
m_{abc_3d}&=A \frac{1}{m!}M^{2m-2}(M^4-2M^2m+(m-1)m)e^{-M^2}
\\
m_{abc_4d}&=A  \frac{1}{m!}M^{2m-2}(-mM^2+m(m-1))e^{-M^2}
\end{align}
with a constant $A=\left|\braket{e_\B k|{\mu}_{e}|g}\right|^{2} \ \left|\braket{\chi_{g,\lambda}^{0}|\mu_p|\chi_{g,\lambda}^1}\right|^2$ that includes the bare phonon and electron dipole matrix elements.
Plugging this into Eq.~(\ref{p4minusp4p}), we obtain
\begin{align}
P_4(t,\tau)+P_4'(t,\tau)&\sim \sum_m \frac{A}{m!}e^{-M^2}M^{2m}(M^2-m)f_{abd}(t,\tau)
\notag\\
&\times\Bigl[\frac{1}{(\omega_{bc_1}+\omega_{2})\sigma_2^2 -(\omega_{c_1d}+\omega_{3})\sigma_3^2}
\notag\\
&+\frac{1}{(\omega_{bc_3}+\omega_{2})\sigma_2^2 -(\omega_{c_3d}+\omega_{3})\sigma_3^2}
+(\sigma_2,\omega_2\leftrightarrow \sigma_3,\omega_3)\Bigr]
\end{align}
with $\omega_{bc_1}=-\omega_{bc_3}=\omega_\lambda$, $\omega_{c_1d}=-(\varepsilon_\B k+ m\omega_\lambda)$, and $\omega_{c_3d}=-[\varepsilon_\B k+ (m-2)\omega_\lambda]$.
The polarization in frequency space given in the method section of the main text  can then be obtained by a Fourier transform
\begin{align}
 P(\omega_t,\omega_\tau)=\int dt d\tau e^{i\omega_t t}e^{i\omega_\tau \tau}[P_4(t,\tau)+P_4'(t,\tau)]]
\end{align}
an by setting $\omega_{da}=\varepsilon_\B k+m\omega_\lambda$, $\omega_{ba}=\omega_\lambda+i\Gamma$, where we have accounted for the broadening $\Gamma$ of the phonon transition (we can ignore the broadening in the interband transition frequency $\omega_{da}$, because the electronic states form a continuum).

\subsection{Toy models of electron-momentum dependent EPC of optical modes}

\subsubsection{Longitudinal optical phonons in the SSH model}

In this section, we derive an estimate for the k-dependence of the EPC strength of optical phonons in a solid using the one-dimensional Su-Schrieffer-Heeger (SSH) model. Our starting point is a one-dimensional tight-binding chain, where the neighboring atoms are connected by springs. The displacement of the atoms $x_j$ alters the hopping strengths of the electrons to the neighboring sites $t_{j,j+1}= t+g(x_j-x_{j+1}) $ giving rise to the SSH Hamiltonian
\begin{align}
 H_{\rm SSH}=&-t\sum_j (c_j^\dag c_{j+1}+{\rm h.c.})+\sum_j \Bigr(\frac{p_j^2}{2m}+\frac{K }{2}(x_j-x_{j+1})^2\Bigl)\notag\\
 &-g\sum_j (x_j-x_{j+1}) (c_j^\dag c_{j+1}+{\rm h.c.}).
\end{align}
Here, $K$ is the spring constant, $m$ the atomic mass, $t$ the hopping strength and $g$ the EPC strength.
We can expand the atomic displacement operators $x_j$ and the corresponding conjugate momenta $p_j$ in a Fourier series using the phonon creation and annhilation operators $b_q^\dag$ and $b_q$, and do the same for the electronic operators $c_j$, and we obtain
\begin{align}
 x_j&=\sum_q \frac{1}{\sqrt{2mN\omega_q}}e^{-iqj}(b^\dag_{-q}+b_q)\\
 p_j&=i\sum_q \sqrt{\frac{m\omega_q}{2N}}e^{-iqj}(b^\dag_{-q}-b_q)\\
 c_j&=\frac{1}{\sqrt{N}}\sum_k e^{-ikj} c_k
\end{align}
The Hamiltonian takes its usual form in momentum representation
\begin{align}
 H_{\rm SSH}=\sum_k \epsilon_kc_k^\dag c_k+\sum_q \omega_qb_q^\dag b_q+H_{\rm EPC}
 \end{align}
with $\omega_q^2=(4K/m)\sin^2(q/2)$. The EPC term can be written as
\begin{align}
H_{\rm EPC}
& =-\frac{1}{N}\sum_j \sum_{kk'q} \frac{g}{\sqrt{2mN\omega_q}}
 e^{-i(k-k'+q)j} (1-e^{-iq})(e^{-ik}+e^{ik'})
 (b^\dag_{-q}+b_q) c_{k'}^\dag   c_{k}\\
& = \sum_{kq} g(k,q)
 (b^\dag_{-q}+b_q) c_{k+q}^\dag   c_{k}
\end{align}
with
\begin{align}
 g(k,q)=\frac{2ig}{\sqrt{2mN\omega_q}}
 [\sin k -\sin (k+q)].
\end{align}
The above model  has only acoustic phonons, but we can also describe optical phonons by assuming an alternating strength of spring constants along the chain, $K\to K_j=K+(-1)^j\delta K$, thereby doubling the real space unit cell. In the reduced zone scheme the $q=\pi$ phonon is mapped to the $q=0$ optical mode. The effective electron-phonon coupling is to zeroth order in $\delta K$ given by the above expression
\begin{align}
  g_{{\rm optical}, q=0}(k)\equiv g(k,q=\pi)=\frac{4ig}{\sqrt{2mN\omega_q}}\sin k .
\end{align}
We conclude that the simplest toy model of an electron-momentum dependent EPC predicts a vanishing coupling strength of a $q=0$ optical mode to the electrons in the center of the electronic Brillouin zone.

\subsubsection{Transverse optical phonons}

While the SSH model directly applies to longitudinal phonons, it does not immediately generalize to transverse phonons. This is because the distance between the atoms only changes quadratically with the transverse atomic displacements and, hence, there would be no linear coupling of the electronic hopping to transverse phonons. Here, we instead employ a one-dimensional toy model with a two atomic basis that couples to transverse phonons in linear order. We will see below that the EPC in this model leads to an effective renormalization of the hopping strength akin to the SSH model.
Our model is a simplification of the model for the A$_{1g}$ and B$_{1g}$ phonons of the CuO planes in the cuprates put forward in \textcolor{blue}{[6, 7]}. 
Specifically, we consider a one-dimensional chain along the $x$-direction with alternating $A$ and $B$ sites, length of unit cell one, and a single electronic orbital on each atom, namely, an $s$-orbital on sublattice $A$ and a $p_x$ orbital on sublattice $B$. Because of the antisymmetry of the $p$-orbital wavefunction the hopping from any $A$ site to the two $B$ sites to the left and right have opposite signs.  We can, therefore, write the purely electronic part of the Hamiltonian of this system as
\begin{align}
 H_{el}=\sum_j\epsilon (c_{A,j}^\dag c_{A,j}-c_{B,j}^\dag c_{B,j})+ t\sum_j (c_{A,j}^\dag c_{B,j}-c_{B,j}^\dag c_{A,j+1}+{\rm h.c.})
\end{align}
where we have chosen zero energy to coincide with the average of the onsite energies of the $A$ and $B$ atoms and we assume $\epsilon>0$.
The Fourier transform of the electronic Hamiltonian reads
\begin{align}
c_{A,j}=&\frac{1}{\sqrt{N}}\sum_k e^{-ikj} c_{A,k}, \qquad
c_{B,j}=\frac{1}{\sqrt{N}}\sum_k e^{-ik(j+1/2)} c_{B,k}\\
 H_{el}=&\sum_k (c^\dag_{A,k},c^\dag_{B,k})
 \begin{pmatrix}
\epsilon & -2it\sin (k/2)\\
2it\sin (k/2) & -\epsilon
 \end{pmatrix}
\binom{c_{A,k}}{c_{B,k}}.
\end{align}
Note that the unit cell has length one and so the $B$ atoms are at the half-integer positions.
The band structure is given by $ E_{k,\pm}=\pm [\epsilon^2+4t^2\sin^2(k/2)]^{1/2}$ and we can project the Hamiltonian to the upper band described by annihilation operator $d_k$ using
\begin{align}
 \binom{c_{k,A}}{c_{k,B}}&\to N_k\binom{\epsilon+E_{k,+}}{2it\sin(k/2)}d_k
\end{align}
with the normalization constant $ N_k^{-2}=(\epsilon+E_{k,+})^2+4t^2\sin^2(k/2)$. We now add phonons to the model, focusing on a single phonon mode with frequency $\omega_{B,q}$ involving transverse displacements of the $B$ atoms while ignoring any motion of the $A$ atoms. In the example of the CuO planes in the cuprates this corresponds to a transverse motion of the O atoms whereas the Cu displacement is zero (B$_{1g}$ phonon) or very small due to the rigidity of out-of-plane bonds (A$_{1g}$ phonon) \textcolor{blue}{[6]}. 
We assume a coupling of these $B$ displacements to the local electron density on the B atoms,
\begin{align}
 H&=H_{el}+\sum_q \omega_{B,q}b_q^\dag b_q+H_{\rm EPC}\\
H_{\rm EPC}
& =- g\sum_j x_j c_{B,j}^\dag c_{B,j} =-\sum_{k,q}\frac{g}{\sqrt{2mN\omega_q}} (b_q+b_{-q}^\dag)  c_{B,k+q}^\dag c_{B,k}.\label{micro_epc}
\end{align}
When we project the EPC to the upper band, we obtain
\begin{align}
H_{\rm EPC}
& =- \sum_{k,q}g(k,q) (b_q+b_{-q}^\dag)  d_{k+q}^\dag d_{k},\label{epc}
\end{align}
with
\begin{align}
 g(k,q)=\frac{4gN_kN_{k+q}t^2}{\sqrt{2mN\omega_q}}\sin(k/2)\sin[(k+q)/2].
\end{align}
For $q=0$, this coupling constant indeed vanishes quadratically at $k=0$. This can be understood by noticing that the electrons at $k=0$ are localized on the $A$ sites and are not directly affected by the phonons, which are localized on the B sites. Electrons at $k\neq 0$ with a finite kinetic energy, however, are partially localized on the $B$ sites due to hopping and therefore couple to the phonons. 

We can even make this intuition that the phonons couple to the kinetic energy of the electrons more explicit by evaluating the real space form of Eq.~(\ref{epc}). For simplicity, we focus on $q=0$ and $k\simeq 0$ so that we can set $N_k\simeq {\rm const.}$. The EPC term takes the form
\begin{align}
H_{{\rm EPC},q=0}
& =- \sum_{k}\alpha (1-\cos k)(b_0+b_{0}^\dag)  d_{k}^\dag d_{k},
\end{align}
with an appropriately defined constant $\alpha$. We perform an inverse Fourier transform and obtain
\begin{align}
H_{{\rm EPC},q=0}
& =- \frac{\alpha}{2}\sum_{i} (b_0+b_{0}^\dag)  (2d_{i}^\dag d_{i}-d_{i}^\dag d_{i+1}-d_{i+1}^\dag d_{i}),
\end{align}
We find that, even though the phonons in the  microscopic model in Eq.~(\ref{micro_epc}) only couple to the electronic density on the $B$ sites, the EPC in the effective model for one particular band modulates the hopping strength.

\newpage
\section{Supplementary Figures}\label{sec2}
\vspace{1cm}
\includegraphics[width=\textwidth]{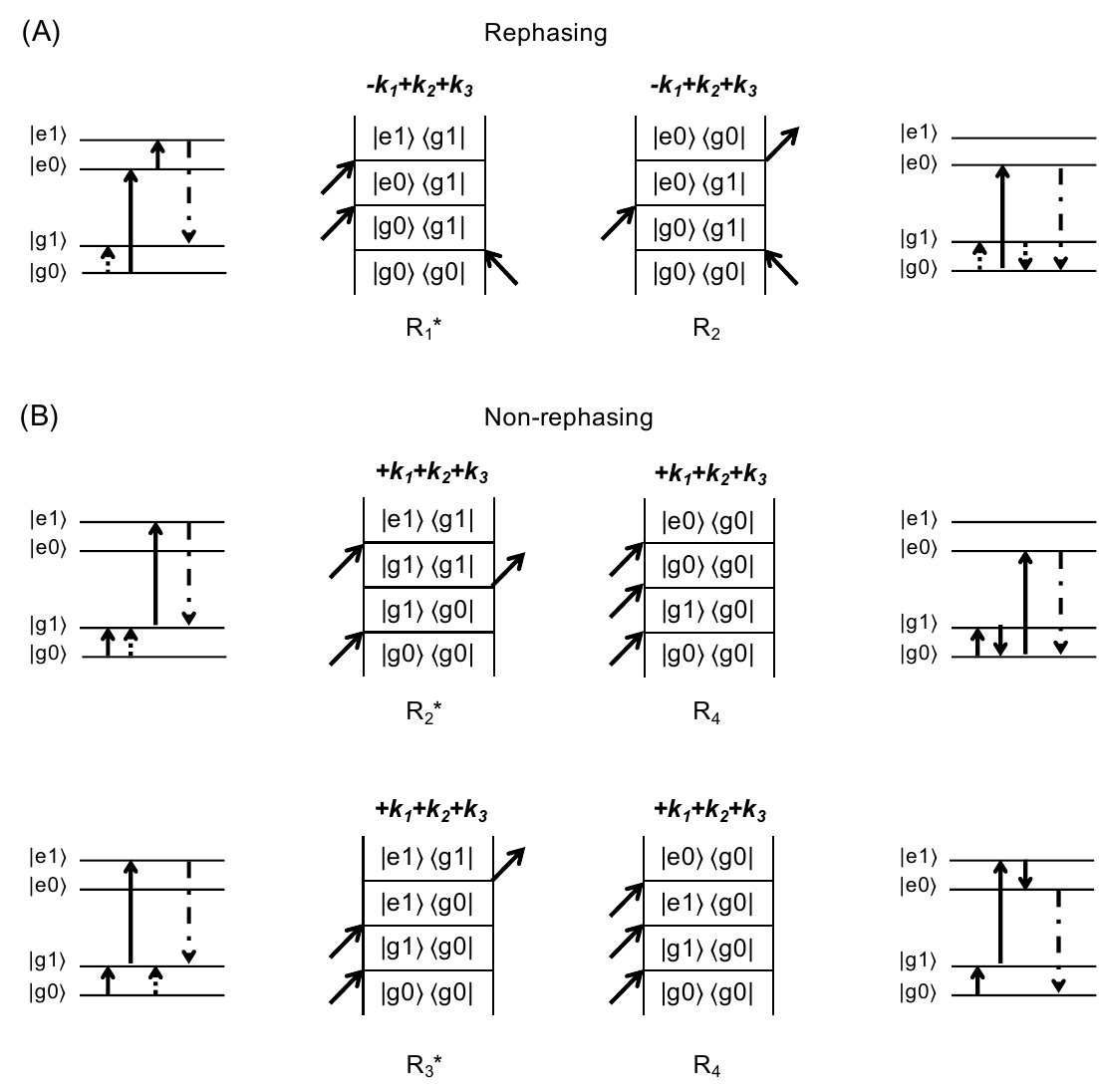}\\
\noindent {\bf Figure S1.} Double-sided Feynman diagrams (middle column) and the corresponding energy level diagrams (left and right columns) for all possible Liouville space pathways for ({\bf A}) rephasing, and ({\bf B}) nonrephasing measurements in the absence of electron-phonon coupling. $\ket{nv}$ indicates the electron-phonon eigenstate at a given moment (the shorthand notation defined to be $\ket{nv}\equiv \ket{n_{\mathbf{k}},\chi_{n,\lambda}^v}$). The vertical solid (dotted) arrow in the ladder diagram denotes action of the external electric field on the ket (bra), and the vertical dashed dot arrow denotes the final emission. The diagonal arrows pointing to the right (left) represents a contribution of the electric field ${\mathbf E_j}e^{-i{\omega}{_j}t+i{\mathbf k{_j}}\cdot{\mathbf r}}$ (${\mathbf E{^*_j}}e^{+i{\omega}{_j}t-i{\mathbf k{_j}}\cdot{\mathbf r}}$) to the polarization, where the frequency $\omega_j>0$.
\vspace{1cm}

\newpage
\vspace{1cm}
\includegraphics[width=\textwidth]{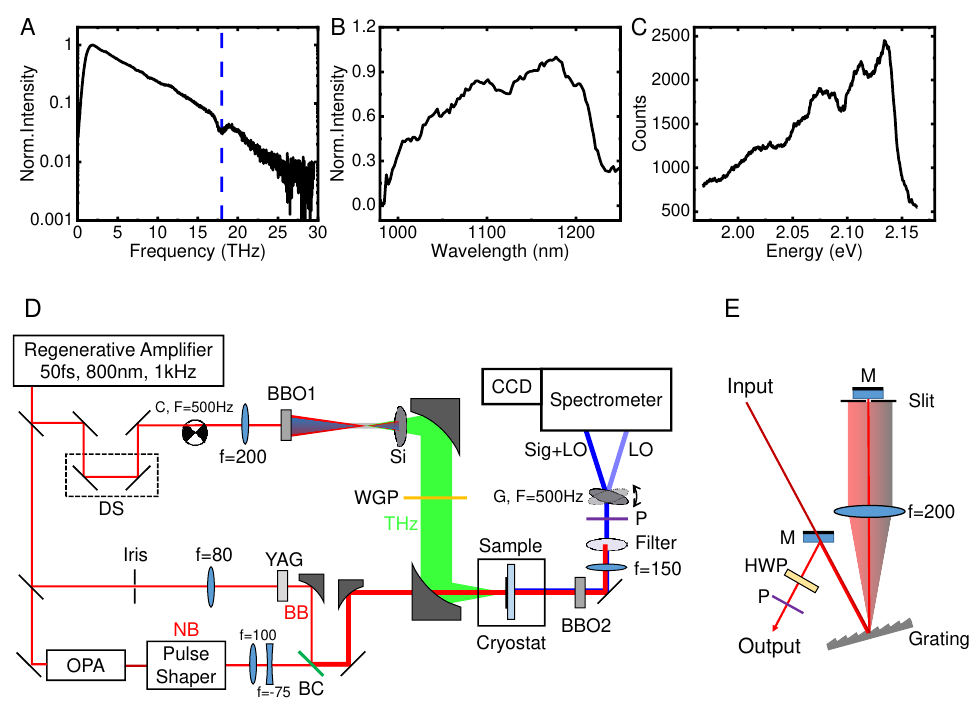}\label{fig_S2}\\
\noindent {\bf Figure S2.} {\bf A,} Intensity of broadband THz pulse. The blue dash line marked the dip at $\sim$18 THz is from the phonon absorption of silicon wafer that is inserted to selectively transmit the generated THz pulses. {\bf B,} Spectrum of BB-NIR measured at sample position generated from YAG crystal. {\bf C,} The spectrum of local oscillator field. {\bf D,} Experimental setup for 2D EPC measurement. DS: delay stage, C: chopper, BC: beam combiner, Si: silicon wafer filter, WGP: wire-grid THz polarizer, P: polarizer, G: mirror galvanometer. {\bf E,} Layout of the Pulse Shaper. M: mirror, HWP: half waveplate. Here, the input beam is from OPA and the output beam is the NB pulse.
\vspace{1cm}

\newpage
\vspace{1cm}
\includegraphics[width=\textwidth]{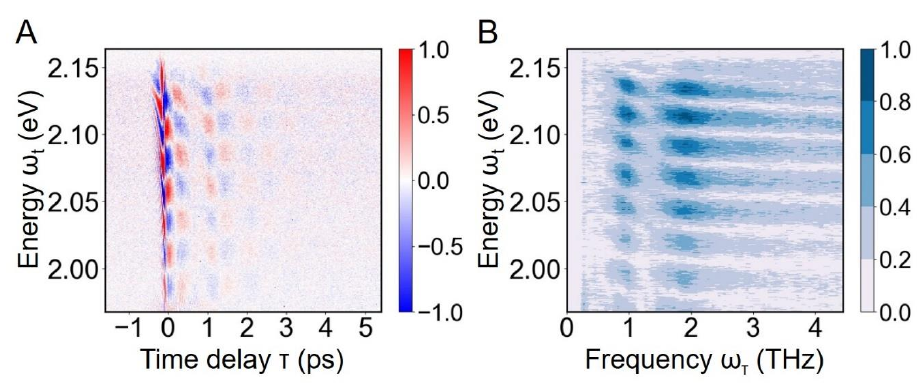}\label{fig_S3}\\
\noindent {\bf Figure S3.} {\bf A,} The measured 2D EPC spectrum, I($\omega_t$, $\tau$) in time domain of MAPI on 0.3 mm glass substrate at room temperature (where the center wavelength of NB = 1350 nm). {\bf B,} Fourier transformed spectrum, I($\omega_t$, $\omega_\tau$) of {\bf A} along time delay.
\vspace{1cm}

\vspace{1cm}
\includegraphics[width=\textwidth]{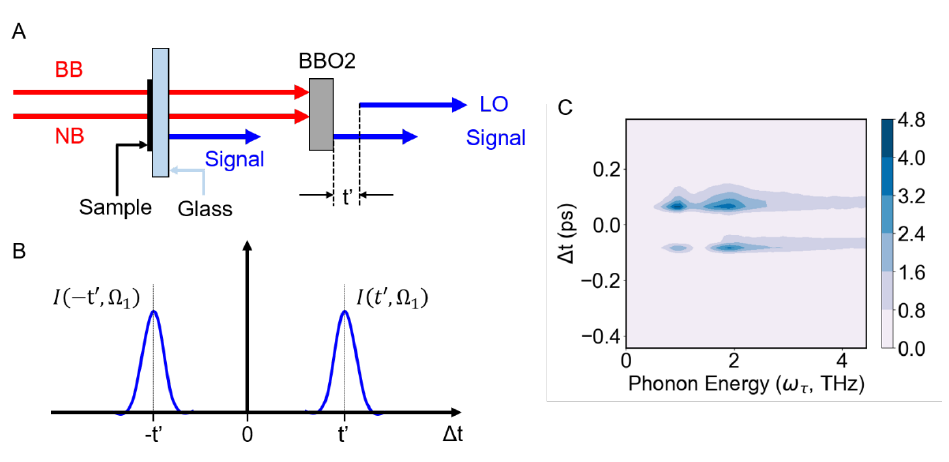}\label{fig_S4}\\
\noindent {\bf Figure S4.} Schematics of quadrant (excitation pathways) separation in 2D EPC spectroscopy. {\bf A,} Pulses configuration for quadrant separation. The glass substrate between the sample and the BBO2 crystal induces a time delay $t’$ between $E_{Signal}$ and $E_{LO}$. The time-separated rephrasing/non-rephasing signal is shown in {\bf B}. {\bf C,} Inverse fast Fourier transform (IFFT) along electron energy axis shown in Figure S3. B. Here the time shift ${\Delta}t=t’\approx 85fs$.
\vspace{1cm}

\newpage
\vspace{1cm}
\includegraphics[width=\textwidth]{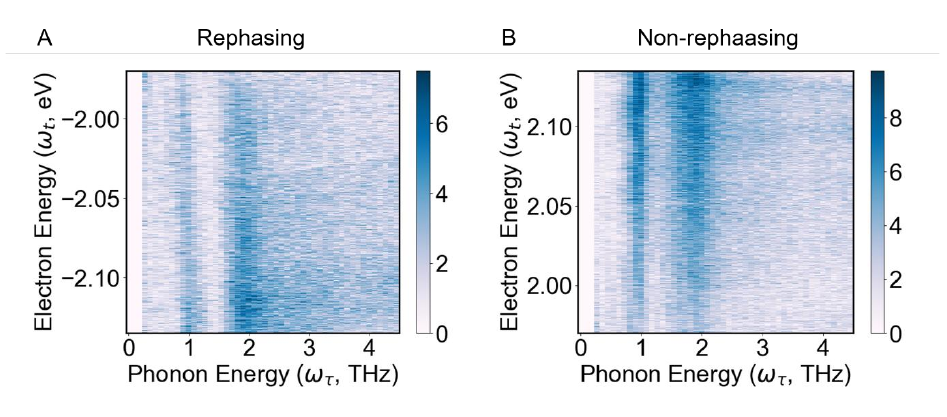}\label{fig_S5}\\
\noindent {\bf Figure S5.} Absolute spectra for {\bf A,} Rephasing and {\bf B,} non-rephasing signals.
\vspace{1cm}

\vspace{1cm}
\includegraphics[width=\textwidth]{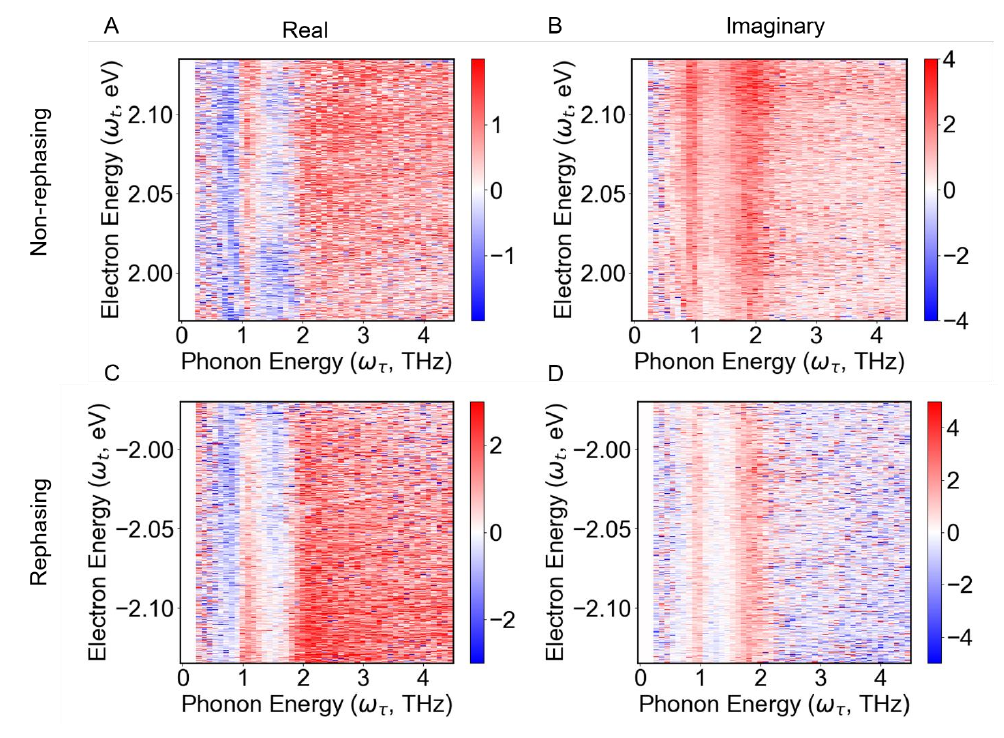}\label{fig_S6}\\
\noindent {\bf Figure S6.} Real ({\bf A, C}) and Imaginary ({\bf B, D}) spectra, for non-rephasing ({\bf A, B}) and rephasing ({\bf C, D}) pathways.
\vspace{1cm}

\vspace{1cm}
\includegraphics[width=\textwidth]{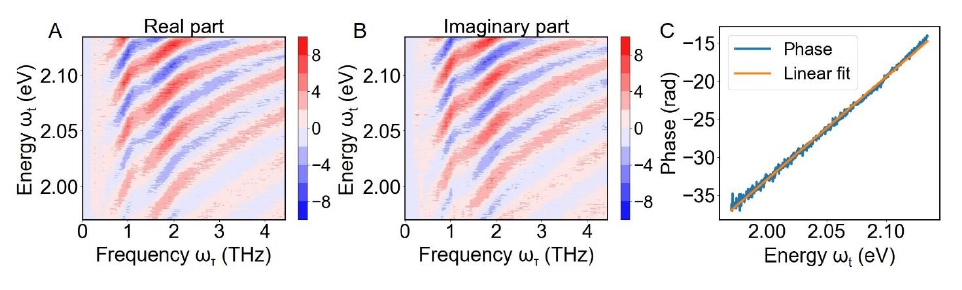}\label{fig_S7}\\
\noindent {\bf Figure S7.} The real {\bf A} and imaginary {\bf B} part of 2D EPC spectrum of MAPI contributed from non-rephasing pathways. {\bf C,} A linear dispersion of the emitted EPC signal phase (blue) with linear fit (orange) at ${\omega}{_\tau} = $ 1 THz of Fig. 2(a).
\vspace{1cm}

\vspace{2cm}
\includegraphics[width=\textwidth]{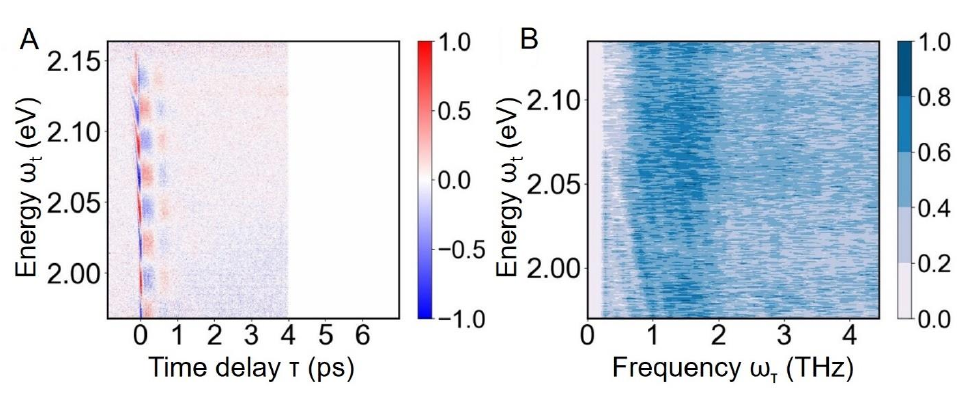}\label{fig_S8}\\
\noindent {\bf Figure S8.} The non-resonant response of SiN$_x$. {\bf A,} The 2D EPC spectrum of SiN$_x$ in time domain measured at room temperature for phasing. Here we use zeros padding the time domain spectrum to keep the Fourier transform window identical to that for the MAPI analysis. {\bf B,} The 2D EPC spectrum of SiN$_x$ in frequency domain contributed from nonrephasing pathways.
\vspace{1cm}

\vspace{1cm}
\includegraphics[width=\textwidth]{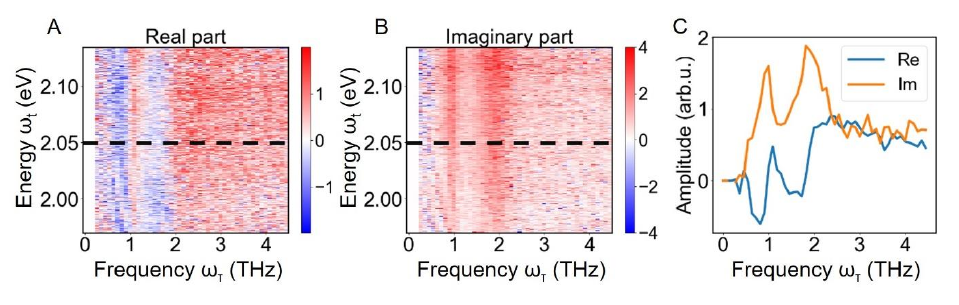}\\
\noindent {\bf Figure S9.} Real, {\bf A,} and imaginary, {\bf B,} part of 2D EPC spectra of MAPI contributed from non-rephasing pathways after phase correction with 2D EPC spectra of SiN$_x$ membrane. {\bf C,} The real and imaginary part along black dash line in {\bf A} and {\bf B} clearly shows the $\sim$1 THz and $\sim$2 THz phonon response.
\vspace{1cm}

\newpage
\vspace{1cm}
\includegraphics[width=\textwidth]{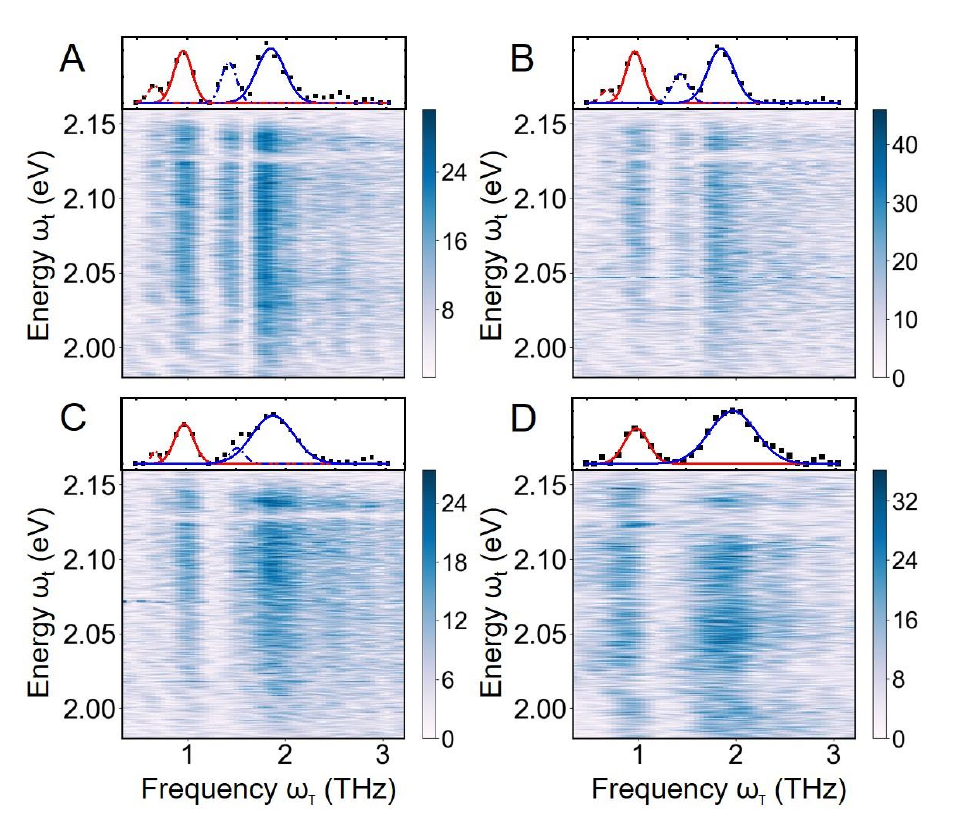}\label{fig_S10}\\
\noindent {\bf Figure S10.} 2D spectra with energy-integrated vibrational EPC response function $S_{EPC}$ (black dot) of MAPI at {\bf A,} 3.7K; {\bf B,} 50K; {\bf C,} 140K and {\bf D,} 200K. Inset: the red dot-dash, red solid, blue dot-dash and blue solid lines are Gaussian fitting of $\sim$0.7 THz, $\sim$1 THz, $\sim$1.5 THz and $\sim$2 THz phonon modes, respectively.
\vspace{1cm}

\newpage
\vspace{1cm}
\includegraphics[width=\textwidth]{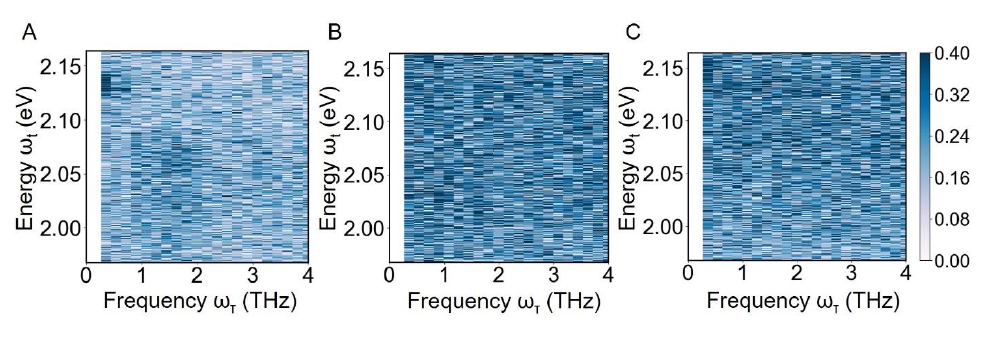}\label{fig_S11}\\
\noindent {\bf Figure S11.} 2D EPC spectra of MAPI in {\bf A,} XYYX; {\bf B,} XXYY and {\bf C,} XYXY polarization configurations.
\vspace{1cm}

\newpage
\vspace{1cm}
    \includegraphics[width=0.95\textwidth]{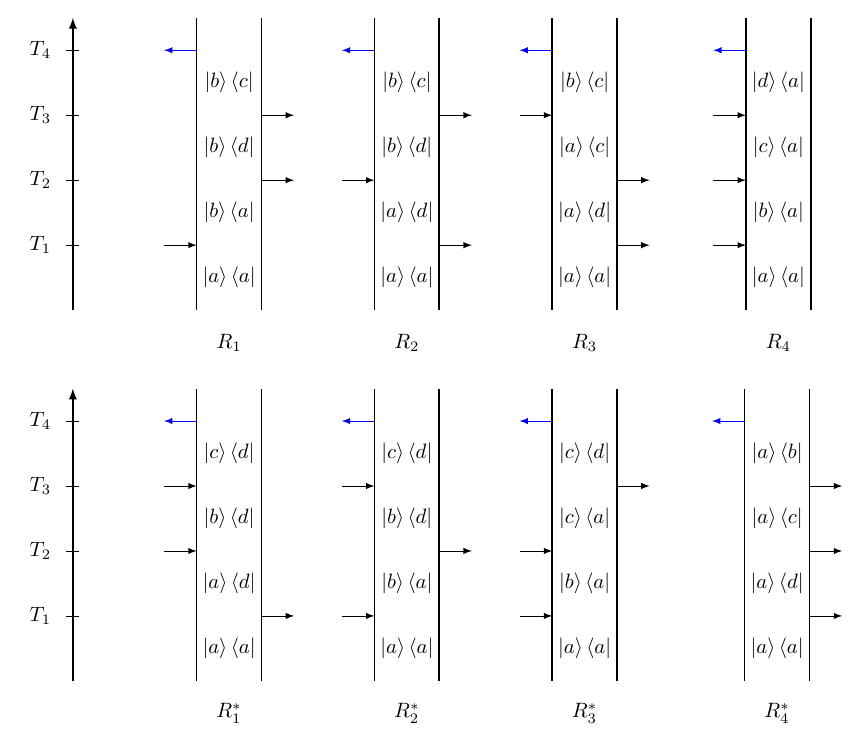}\label{fig_S12}\\
\noindent {\bf Figure S12.} Feynman Diagrams for the polarization. The contribution from $R_{1/4}^*$ and $R_{2/3}$ vanishes because they require an emission from the ground state. The $R_1$ diagram is also negligible because it requires emission from the ground state from both second and third pulse. Therefore, only $R_{2/3}^*$ and $R_4$ contribute to the polarization. Note that $R_{2/3}^*$ are nonzero because the second and third pulse overlap with each other.
\vspace{1cm}

\backmatter
\includepdf{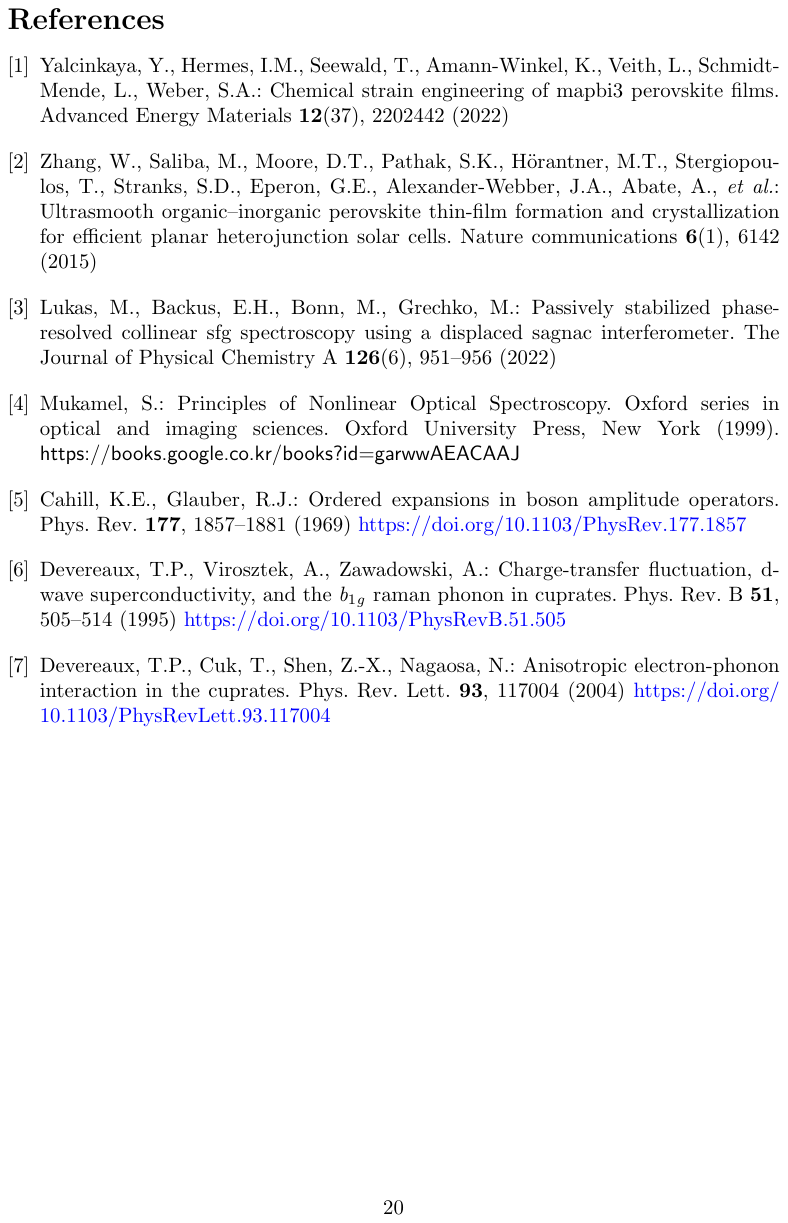}

\end{document}